\documentclass[journal]{IEEEtran}
\newcommand{\nop}[1]{}
\usepackage{xcolor}
\usepackage{amssymb}
\usepackage{algorithm}
\usepackage[noend]{algorithmic}
\usepackage{amsmath}
\usepackage{amssymb}
\usepackage{array}
\usepackage{graphicx}
\usepackage{pifont}
\usepackage{epsfig,tabularx,amssymb,amsmath,subfigure,multirow,booktabs,epstopdf}
\usepackage{balance}
\usepackage{mathptmx}
\usepackage{cite}
\usepackage{url}

\newtheorem{definition}{Definition}

\newtheorem{theorem}{Theorem}

\ifCLASSINFOpdf
\else
\fi
\allowdisplaybreaks[4]

\begin{document}
%\title{Optimizing Request Scheduling of Hybrid\\Edge Service in Edge Computing}
\title{HyEdge: Optimal Request Scheduling in Hybrid Edge Computing Environment}
\author{Siyuan~Gu\IEEEauthorrefmark{1},
        Deke~Guo\IEEEauthorrefmark{1}\IEEEauthorrefmark{2},~\IEEEmembership{Senior Member,~IEEE,}
       Guoming~Tang\IEEEauthorrefmark{1},
       Yuchen~Sun\IEEEauthorrefmark{1}
       ~and Xueshan~Luo\IEEEauthorrefmark{1}
\IEEEcompsocitemizethanks{\IEEEcompsocthanksitem
\IEEEauthorrefmark{1}Science and Technology on Information Systems Engineering Laboratory
National University of Defense Technology, Changsha Hunan {\rm410073}, P.R. China.\protect
\IEEEauthorrefmark{2}College of Intelligence and Computing, Tianjin University, Tianjin {\rm 300350}, P.R. China.\protect
E-mail: gusiyuan18@nudt.edu.cn, guodeke@gmail.com, gmtang@nudt.edu.cn, sunyuchen18@nudt.edu.cn, xsluo@nudt.edu.cn.}
}
\maketitle\renewcommand{\thefootnote}{\fnsymbol{footnote}}
\footnotetext{Deke Guo is the corresponding author.}

% <-this % stops a space
%\thanks{This work is partially supported by the National Natural Science Foundation for Outstanding Excellent young scholars of China under Grant No.61422214, National Natural Science Foundation of China under Grant No.61772544, National Basic Research Program (973 program) under Grant No.2014CB347800, the Hunan Provincial Natural Science Fund for Distinguished Young Scholars under Grant No.2016JJ1002, and the Guangxi Cooperative Innovation Center of cloud computing and Big Data under Grant Nos.YD16507 and YD17X11.}
%\thanks{G. Cheng, D. Guo, J. Shi, and Y. Qin are with the National Key Laboratory of Science and Technology on C^4ISR Technology, National University of Defense Technology, Changsha (410005), China (Email: chenggeyao13@nudt.edu.cn; guodeke@gmail.com; jshi1980@163.com; qinyudong12@nudt.edu.cn).}% <-this % stops a space
%}
\markboth{Journal of \LaTeX\ Class Files,~Vol.~?, No.~?, ?~20??}%
{Shell \MakeLowercase{\textit{et al.}}: Bare Demo of IEEEtran.cls for Computer Society Journals}

\maketitle

\begin{abstract}
With the widespread use of Internet of Things (IoT) devices and the arrival of the 5G era, edge computing has become an attractive paradigm to serve end-users and provide better QoS. Many efforts have been done to provision some merging public network services at the edge. We reveal that it is very common that specific users call for private and isolated edge services to preserve data privacy and enable other security intentions. However, it still remains open to fulfill such kind of mixed requests in edge computing. In this paper, we propose the framework of hybrid edge computing to offer both public and private edge services systematically. To fully exploit the benefits of this novel framework, we define the problem of optimal request scheduling over a given placement solution of hybrid edge servers, so as to minimize the response delay. This problem is further modeled as a mixed integer non-linear problem (MINLP), which is typically NP-hard. Accordingly, we propose the partition-based optimization method, which can efficiently solve this NP-hard problem via the problem decomposition and the branch and bound strategies. We finally conduct extensive evaluations with a real-world dataset to measure the performance of our methods. The results indicate that the proposed method achieves elegant performance with low computation complexity.
\end{abstract}


\IEEEpeerreviewmaketitle

\section{Introduction}\label{Intro}
In the past decade, cloud computing has successfully delivered a lot of cloud services to many end-user devices via large-scale data centers. Cloud computing leverages the vast resources available in the data centers to deliver elastic computing power and storage. However, the emergence of ultra-low latency, high-bandwidth consumption network services poses great challenges to the cloud computing model~\cite{Armbrust2013A }. For example, cloud computing has encountered tremendous obstacles to serve the increasing requirements of AR (augmented reality)~\cite{Dunleavy2014Augmented}, Intelligent video acceleration~\cite{Zhang2017Video}, autonomous driving~\cite{Urmson2008Autonomous} applications and beyond.

To tackle such challenges, \emph{edge computing} has been proposed to realize a series of computing environment at the network edge, which is more closer to end-users. Edge computing motivates to serve end-users with a better quality of service (QoS) than cloud computing, e.g., low transmission cost, short response latency, privacy preservation. It has become an attractive paradigm to provision computing resources and services to end users.

Aided by the edge computing, most cloud services (including the aforementioned new network services) can be offloaded to the edge servers for better QoS. These services usually can be accessed by the public (e.g., via open APIs) and serve a large group of users (e.g., in the location-based services). In this paper, we call such services \emph{public edge services}. Currently, the public edge service is the most common case among existing edge services.

\begin{figure}[t]
   \centering
   \includegraphics[width=0.95\linewidth]{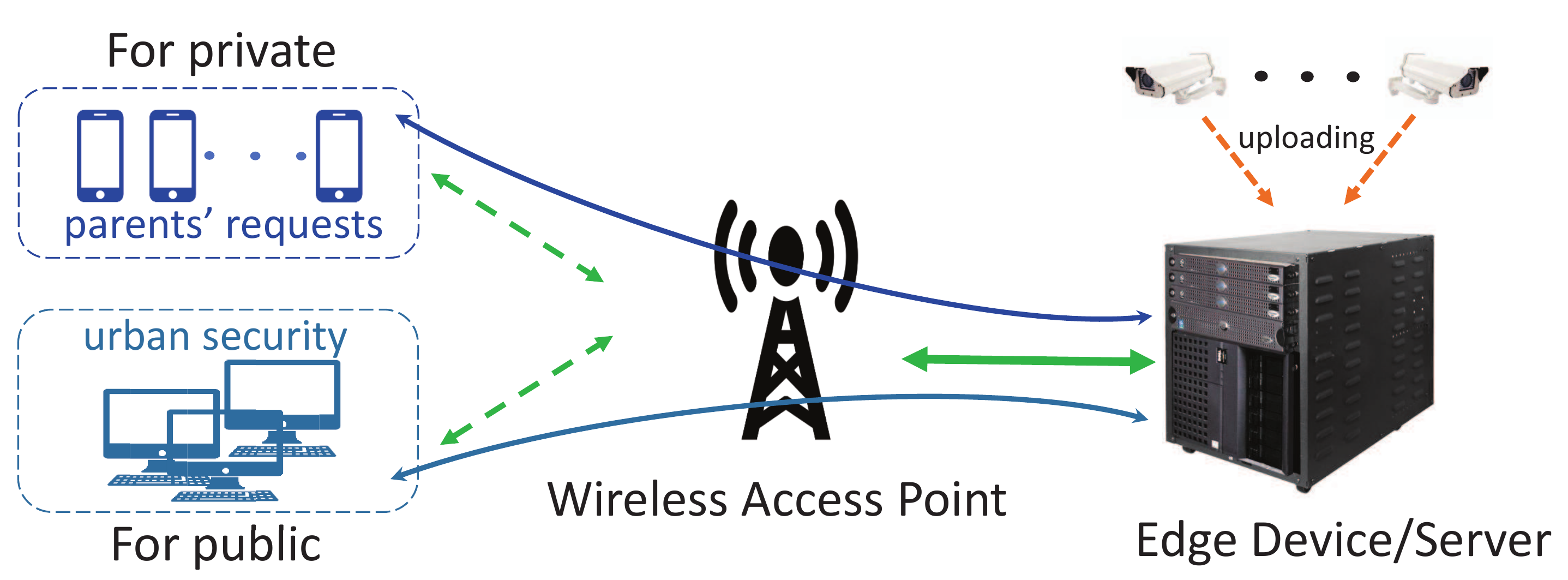}
   \caption{Example of public and private (hybrid) edge services in a kindergarten surveillance scenario.}
 \label{fig-example}
\end{figure}

However, \emph{public edge services} alone cannot meet more elastic and customized needs, such as further improving service performance, extending stable uptime and ensuring high-level security/privacy. Thus, many service providers seek to deploy private services at the edge to make up for it (refer to the following example along with Fig.~\ref{fig-example}). Also, massive personal data, previously managed in private IoT devices, are expected to be transferred to edge servers for centralized processing (to deal with the lack of computation and storage capacity at IoT devices). All the above scenarios call for specialized edge services that tailor for customized needs. We name them \emph{private edge services} in this paper.

\textbf{Example:} Fig.~\ref{fig-example} gives a simple yet concrete example to show the public as well as private services in edge computing. We consider a kindergarten surveillance system with many security cameras installed. An edge server is responsible for storing and processing images and video streams uploaded by these cameras. The cameras are placed in different places for specific purposes, e.g., cameras for parents to observe their children are mainly placed indoors while cameras for urban safety system and street view system are mainly installed outdoors. Under such a situation, the video streams uploaded by different cameras will be stored by the shared infrastructure, and processed and returned according to customers' needs. In order to achieve the above application requirements, the shared edge server should meet the needs of both public  and private customers.

With the rapid emergence of private applications (services), however, most of them are characterized by fast generation and low resource (w.r.t., communication, storage and computation resources) consumption. These features enable the flexibility of implementing the private edge services at the edge. Certainly, assigning an exclusive server to each private edge service is not advisable due to its unacceptable hardware investment. The service providers prefer to leveraging the available resource in an edge server which is mainly configured for public edge services to accommodate some private services. As a consequence, the public and private edge services will coexist on a common edge server, resulting in a \emph{hybrid edge service} environment.

The \emph{hybrid edge service} environment further leads to the complicated \emph{hybrid edge computing}, which differs from the \emph{hybrid cloud computing}~\cite{Linthicum2016Emerging} significantly. First, in hybrid edge computing, the mobility and ``plug-in" style of customer services make profound impacts to the service placement and request scheduling strategies; whereas the cloud is insensitive to and not restricted by such mobility and uncertainty. Thus, it is tricky to embed private services among the public ones, due to the uncertainty caused by user mobility. Second, compared to the hybrid cloud computing, the hybrid edge computing is more efficient at providing delay-sensitive services as it pushes the resources closer to the customers. Nevertheless, when both public and private service requests arrive at the edge server concurrently, how to schedule them is still a new problem especially for large-scale concurrent requests. Third, unlike the abundant resources in the cloud, the resources at the edge are much scarce. Consequently, it is challenging to provide QoS guarantee in the hybrid edge computing environment, with the constraints of available resources provided by the edge servers.

In this paper, we address the above challenges in a systematic way, and summarize them as follows.
\begin{itemize}
\item We formally define the concept and establish the framework of hybrid edge computing, where hybrid edge servers are deployed at the network edge. This framework is a valuable reference for the edge service providers and is helpful to expand edge services with general applicability.
\item We formulate the problem of optimal request scheduling over hybrid edge server placement. To find the optimal request scheduling path (the one minimizes the delay while maximizing the number of requests), we model the problem as a mixed integer non-linear problem (MINLP).
\item We develop the partition-based algorithm for optimal request scheduling of hybrid edge services. The scheduling algorithm carefully transforms the original MINLP problem into four easier sub-problems according to different resource constraints. Then these sub-problems can be solved effectively with the branch and bound method. The results demonstrate that the optimal scheduling solution can be achieved in a quite short time.
\item We also analyze the performance of our algorithm and the effect of edge server settings in hybrid edge computing. The experimental results show that our algorithm and edge server setting strategy are quite efficient and exhibit good scalability with respective to the network size.
\end{itemize}

\nop{
Aiming at tackling these challenges, our contributions can be summarized mainly as follows:

\begin{itemize}
  \item We propose the framework of hybrid edge computing where we deploy hybrid edge servers at the edge of the network to provide hybrid edge services. Our framework is meaningful for service providers to branch out into the edge service with general applicability.
  \item In order to obtain the request scheduling path that minimizes the delay while maximizing the number of requests, we formulate the problem of \emph{optimal request scheduling over hybrid edge server placement} as a mixed integer linear problem(MILP).
  \item Our solution completely divides the original problem into four sub-problems, and find that the branch and bound method is very effective in solving these sub-problems. And the results of our solution demonstrate that optimal scheduling can be archived in a quite short time.
  \item We also analyze the performance of our algorithm and the effect of edge server settings in hybrid edge computing. The experimental results show that our algorithm and edge server settings are very efficient and exhibit very good scalability to the network scale.
\end{itemize}
}

The remainder of this paper is organized as follows. In Section~\ref{relatedwork}, we give a review of related work. In Section~\ref{framework}, we present the formal introduction of hybrid edge computing. Then we propose our problem, analyze the hardness of it, and formulate the problem mathematically in Section~\ref{problem}. We propose our corresponding algorithm with the analysis in Section~\ref{method}. In Section~\ref{evaluation}, we evaluate our algorithm with simulations over a real-world dataset. We conclude this paper in Section~\ref{conclusionfuturework}.

\section{Related Work}\label{relatedwork}

\textbf{Overview of edge computing.} With edge computing becoming popular nowadays, it has become a hot topic to transfer the data processing from the core server to the IoT devices or edge servers and respond the user requests directly at the edge. Bonomi et al.~\cite{Bonomi2012Fog} introduced the concept of fog computing (a.k.a edge computing), analyzed the transition from cloud computing to edge computing and showed its crucial role in the IoT. Shi et al.~\cite{Shi2016The} illustrated the concept of edge computing from a comprehensive perspectives, and also took major potential concerns into consideration, e.g., bandwidth consumption, limited battery life for mobile devices, and security. Specific techniques of edge computing were also braodly investigated, such as variant models (e.g., mobile edge computing (MEC)~\cite{Xu2016Efficient,Mao2017A} and cloudlet~\cite{Jararweh2014Scalable}), objective optimization (e.g., minimum delay~\cite{Liu2016Delay,Wang2017Joint}, QoS improving~\cite{Skarlat2017Towards}), and workload offloading (e.g., workload management~\cite{Habak2017Workload}, workload allocation~\cite{Deng2015Towards}). Although a comprehensive study has been made related to the edge computing technologies, they ignore a key aspect, that is, the in-depth discussion of service types. At present, existing solutions~\cite{Shi2016The,Mao2017A,Skarlat2017Towards} to edge computing only tackle those public services (that can be accessed without limitations of time, location or identification), from a data center cloud to the edge servers in close proximity to the users. As we have mentioned, with the increasing demands for private services and the successive construction of private (edge) clouds, it is an emergency to investigate how those challenges involving hybrid edge computing can be solved, which has not been done yet.

\textbf{Classic placement problem.} A broad spectrum of researches on the ``placement problem''  have been done in the edge computing domain, including application placement~\cite{Wang2017Online,Karagiannis2017Network}, virtual machine(VM) placement~\cite{Wei2017Virtual}, server placement~\cite{Ahuja2012Algorithms,Wang2018Edge}, and service placement~\cite{Skarlat2017Towards,Wang2017Dynamic}. Many network optimization goals (e.g., cost reduction or performance improvement) can be achieved through such placement strategies. Lin et al.~\cite{Lin2018Service} minimized the total cost of placing service entities for social virtual reality applications. Ting et al.~\cite{He2018It} maximized the number of requests processed by joint research on service placement and request scheduling with sharable and non-sharable resources in mobile edge computing. Note that the placement problems often involve storage resources issues since services or contents need to be distributed in servers or VMs. Nevertheless, there were few placement problems relevant to the hybrid services studied in this work.

\textbf{Request scheduling strategy.} When resources are non-sharable, which means there may exist resource competitions between the users, it is necessary to strategically schedule requests or allocate workloads to ensure better QoS, and ultimately achieve a reasonable scheduling performance. A lot of work has been done to offload workloads to the edge server, usually from the data centers~\cite{Ismail2017Towards,Vakilinia2017Energy}. Mao et.al.~\cite{Mao2016Dynamic} greatly improved the quality of computation experience, e.g., the execution latency, by offloading the computationally intensive workloads to the edge servers. Chen et.al.~\cite{ChenDynamic} further discussed the dynamic service request scheduling problem in edge computing to minimize the scheduling cost. Unfortunately, no related work has been done on the scheduling strategies of hybrid service requests.

Overall, to the best of our knowledge, we are the first to put forward the hybrid edge computing and study the problem of optimal service request scheduling, over the hybrid edge server placement. Meanwhile, we also consider the communication and computation resources constraints on edge servers and target at providing effective and efficient solutions to the optimal request scheduling problem.

\section{Framework of Hybrid Edge Computing}\label{framework}
In this section, we first state the concepts of public edge service, private edge service and hybrid edge service, whose descriptions are briefly mentioned in Section~\ref{Intro} and exact definitions have not been found in other literature. We also give the concept of the hybrid edge server which is extremely important for deploying hybrid edge services,etc. Then we will introduce the network model in the Wireless Metropolitan Area Network (WMAN) consisting of many Access Points (APs). Next, we state the mechanism of service request and response and expose the problems faced in the concurrency of hybrid services.

\subsection{Definitions of Edge Services}
The development of edge computing has brought about a brand new distributed computing paradigm, with many new services emerging but not be accurately defined. In our framework, three concepts are presented step by step.

\begin{definition}[Public Edge Service]
Public edge service is the function or the service mainly provided by public sectors or the service providers for meeting massive public network service requirements. It leverages limited public memory and computing resources on devices at the edge in response to the request of users.
\end{definition}

\begin{definition}[Private Edge Service]
Private edge service relies on private resources at the edge to provide customized functionalities, isolated performance, privacy-preserving, and security guarantee to its users. It differs from the public edge service mainly in terms of private deployment and customization features.
\end{definition}

\begin{definition}[Hybrid Edge Service]
Inspired by the previous practice of edge computing and hybrid cloud, hybrid edge service integrates the public and private edge services at the edge with a specific hybrid service placement strategy. A natural problem in this kind of edge service is that whether the efficient network operation and QoS can still be guaranteed or improved?
\end{definition}

So far, we have showed that the task of deploying hybrid edge services at the edge is imminent and critical.

\begin{figure}[t]
   \centering
      \includegraphics[width=3.5 in,scale=1.0]{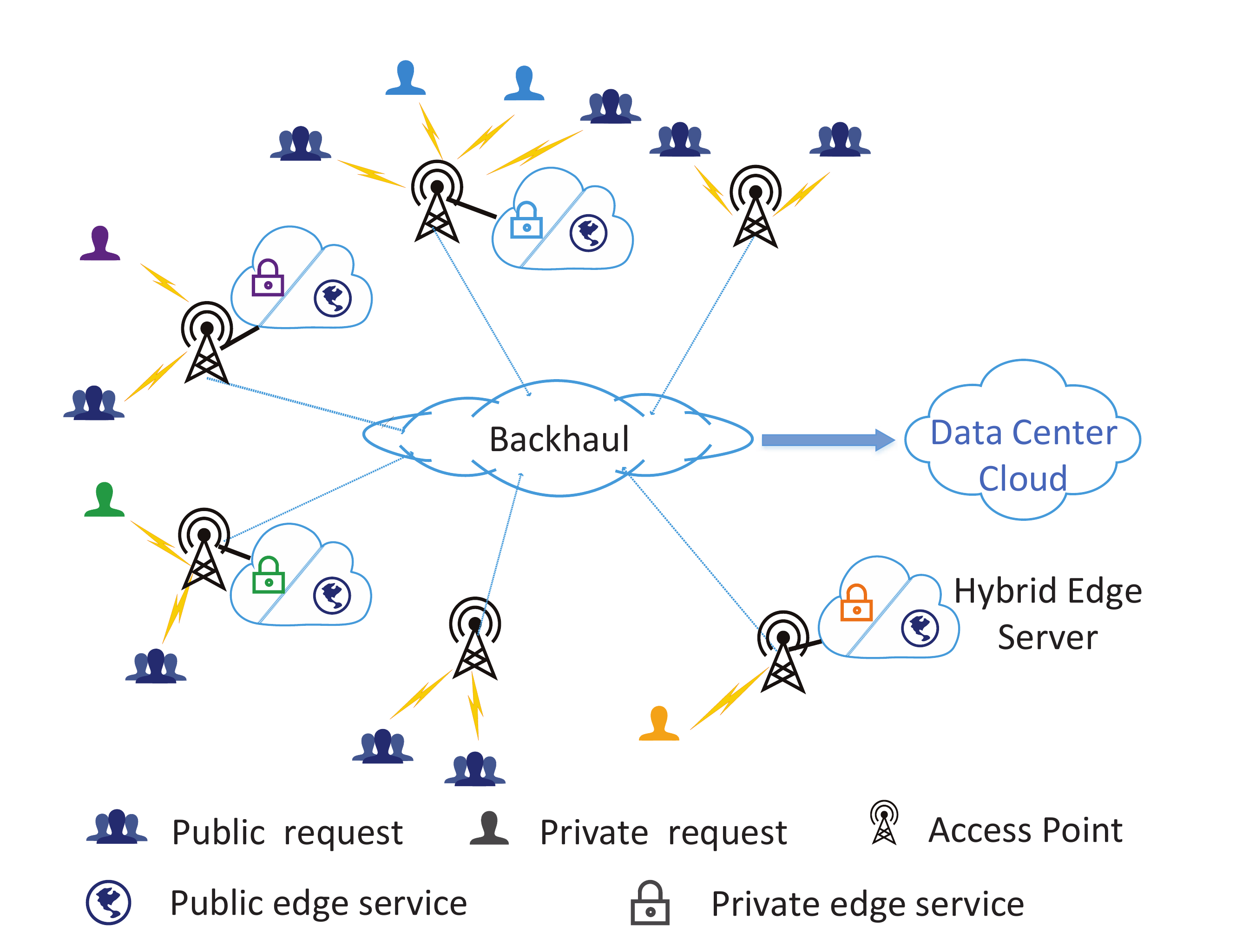}\vspace{-0.1in}
 \caption{A framework of hybrid edge computing.}
 \label{fig-Network}
\end{figure}

\subsection{Network Model}
Our network model is formed in a WMAN environment. In order to provide better services, edge servers, which is much closer to the users and data sources, are implemented as powerful complements to the cloud data centers. For instance, the computation units in the edge servers can be used to preprocess the data before forwarding to the data centers. In this way, the edge servers speed up the data processing, shorten the request response time, and lessen the burden of data centers in the remote cloud.

As shown in Fig.~\ref{fig-Network}, we abstract the network topology as an undirected graph $G (V, E)$, where $V$ is the set of $n$ Access Points (APs), and $E$ is the set of all communication links which interconnect the APs. We use the backhaul network to represent all the links between APs and the remote cloud data center. Let $m$ be the cardinality of $E$. We suppose that we have an average communication delay between an AP $\upsilon_{i}$ and its local service requests.

We deploy the hybrid edge server at the edge of the network to provide collaborative processing with the cloud data center. Such a hybrid edge server has the following properties. First, a hybrid edge server is much smaller than a general hybrid cloud. It has proper flexibility and scalability just like Mobile Edge Cloud (MEC) and is co-located with a wireless access point. Second, there are two kinds of services resided in one hybrid edge server, including the private services, which serve the authenticated requests locally, and the public services which handle the public user requests from the entire edge environment. Third, unlike the communication resource which is sharable among AP, the computation resource is non-sharable in a hybrid edge server. The reason is that there is a strict resources partition in edge cloud to handle different kinds of service requests (i.e., private service requests and public service requests).

Besides, the capacity of each hybrid edge server may be different and adjustable. Likewise, each service may declare heterogeneous requirements to storage, communication and computation resource. In this paper, for simplicity, we assume that each hybrid edge server has a uniform computation capacity and adequate storage space and each AP has a uniform communication capacity.

\subsection{Service Request and Response}
In the framework of hybrid edge computing, the computing power of the data center can be supplemented by utilizing resources on the edge-to-cloud network paths to cope with the increasing number of hybrid service requests and reduce the load on the data center cloud.

The traditional request/response mechanism in cloud computing has to transmit the request data to the cloud data center. Only after the completion of data processing, the generated results will be returned to the client. The unpredictable geographical locations of the service request further aggravate the response delay. This is definitely unacceptable for delay-sensitive applications (e.g., early warning systems, self-driving vehicles, etc.). Consequently, the cloud computing framework is not competent in such delay-sensitive applications. However, with the help of edge servers, hybrid edge computing can complete data processing and respond to the requests much faster, such that delay-sensitive applications can also be supported with better QoS.

In this paper, we collectively refer to a user's request as a service request. The service requests are categorized as the public request and private request, which demand public and private functions on hybrid edge servers, respectively. In this homogeneous network model, we assume that service requests can be predicted for a certain time period and the ratio between the number of public requests and the number of private requests is range in a given interval for any AP.

When a user request is sent to a local AP, it will be processed locally if a co-located hybrid edge server has sufficient available capacity. Otherwise, this request will be directed to a competent neighboring server or the cloud data center. We assume that all service requests need the same computation resources. Similarly, the service request also needs to meet the communication resource constraints on the paths during the transmission process. In order to maximize the number of hybrid service requests, we must derive out the scheduling strategy with limited resources. Before that, we first describe the resources constraints in Section~\ref{problem}.

\begin{table}[t]
    \caption{SUMMARY OF NOTATIONS}\label{table}
    \begin{center}
        \begin{tabular*}{8.8cm}{@{\extracolsep{\fill}}ll}
	\hline
	    \emph{Notation}        & \emph{Description}  \\
		\hline
    	$G$             & graph of communication network topology \\
    	$V$             & the vertices (APs) set of $G$ \\
    	$E$             & the edges (links) set of $G$ \\
    	$K$             & the computation capacity of a hybrid edge server \\
    	$W$             & the communication capacity of a hybrid edge server \\
        $\lambda$       & the communication delay between any hybrid edge server \\
                        & and the data center cloud \\
        \hline
        $m$             & total number of hybrid edge servers \\
        $n$             & total number of Access Points \\
    	$i/j$           & index of Access Points \\
    	$\alpha$        & the ratio of private resources in a hybrid edge server \\
    	$\beta$         & the ratio of private requests at an Access Point \\
    	\hline
    	$\upsilon_{i}$  & the $i$th Access Point \\
    	$\pi_{i}$       & the fixed communication delay between AP $\upsilon_{i}$ and its \\
                        & local user requests \\
    	$\theta_{i}$    & the number of user requests at AP $\upsilon_{i}$ \\
    	$x_{i}$         & indicator value about if there is a hybrid edge server at AP $\upsilon_{i}$ \\
    	$\zeta_{i}$     & the number of offloaded requests from AP $\upsilon_{i}$ to the data \\
                        & center cloud \\
    	$\chi_{i}$      & the number of requests that will be communicated locally \\
                        & at AP $\upsilon_{i}$ \\
    	$\xi_{i,j}$     & the communication delay when the requests at AP $\upsilon_{i}$ are \\
                        & scheduled to the AP $\upsilon_{j}$ \\
    	$y_{i,j}$       & the number of requests which are routed from AP $\upsilon_{i}$ to $\upsilon_{j}$ \\
                        & co-located with the hybrid edge server \\
    	\hline
        \end{tabular*}
    \end{center}
\end{table}

\section{Problem Formulation}\label{problem}

In this section, we first make a preliminary analysis and introduce main constraints for further discussion. Then we propose the problem of optimal request scheduling over hybrid edge server placement and give our formulation. Finally, we will prove the hardness of our problem. For ease of reference, all notations used in this paper are listed in TABLE ~\ref{table}.

\subsection{Preliminary analysis}
When a private service is implemented in a hybrid edge server, its users can access it more conveniently. Meanwhile, the cloud data center also delivers replicas of public services on the hybrid edge servers. In both cases, each hybrid edge server has enough storage space, since the corresponding functions of both public and private services are successfully set up in the server. Therefore, storage resource competition does not need to be considered in this paper.

Since the hybrid edge server has abundant memory to reside hybrid services, we only consider the limitation of communication and computation capacity in our hybrid edge computing model. Besides, the competition will occur when the private and public edge requests are launched simultaneously. Thus, we start with a brief introduction to resource constraints.

\subsubsection{Communication Constraints of Local Request Scheduling}
APs only receive requests from its covered users with respect to the communication constraints. At the user end, its communication resource is duplexed by both private and public requests. We will only consider the local communication constraints problem since communication capacity of inter-AP links is rich enough. Let $W$ be the number of requests which can be communicated locally at one AP in a time slice. We divide the user request into the public request and the private request. Let $\beta$ be the ratio of private requests to total requests at any AP; thereby $1-\beta$ is the ratio of public requests to total requests at any AP. Generally, the number of requests handled by a local AP can never be larger than the value of $W$. Therefore, we let $\theta_{i}$ be the number of total user requests at AP $\upsilon_{i}$ and $\chi_{i}$ be the number of requests that communicate directly with $\upsilon_{i}$. Then we have $\chi_{i}\mathrm{=}\min \{\theta_{i}, W\}$. When the requests are processed locally, we will denote the fixed delay as $\pi_{i}$, which means the communication delay from AP $\upsilon_{i}$ to its co-located hybrid edge server. Furthermore, if the requests at AP $\upsilon_{i}$ are scheduled to another AP $\upsilon_{j}$, the communication delay will be quantified as $\xi_{i,j}$.
\subsubsection{Computation Constraints of Local and Remote Request Scheduling}\label{ps}
Let $K$ be the uniform computation capacity of any given hybrid edge server. That is, $K$ quantifies the maximum number of requests that the server can process and respond. We further divide the computation capacity into the private part and public part, which represents the maximum number of private and public requests that the server can process respectively. Let $\alpha$ is the ratio of private part, then $1-\alpha$ be the ratio of the public part. Normally, private part of any given hybrid edge server can process all the private requests in a time slice since we prefer to meeting the requests of private users first. However, for the public part, since local computation resources cannot always meet their demands, thus the unsatisfied part of public requests will be scheduled to other APs co-located whose corresponding hybrid edge servers have extra computing capacity. Besides, the hybrid edge server may have to redirect some requests to the remote data center because of its limited computing capacity. We denote the number of redirected requests from AP $\upsilon_{i}$ to the data center as $\zeta_{i}$. A by default assumption for the data center is that it has abundant resources to process all received requests. The delay between the hybrid edge server and the data center is a constant  $\lambda$.

\subsection{Optimal Request Scheduling over Hybrid Edge Server Placement (ORS-HESP)}
Before scheduling the request of hybrid edge service, we must seriously consider the placement of the hybrid edge server and resources assignment. Hybrid edge servers can be used to accommodate public and (or) private edge services (data, algorithms, etc.). But the private services should be deployed close to private requests to shorten the response delay. Therefore it is reasonable to co-locate the hybrid edge server at the AP which covers the corresponding private user requests. Certainly, resources should be dynamically and adaptively adjusted according to the real-time number of service requests.

We mainly study the request scheduling under resource allocation during a time window, so that the number of requests is predictable. Then we study how to schedule our user requests to certain edge servers so that the total delay is minimized with respect to the computation and communication resource constraints.
Since we have $m$ APs, we will introduce the set of variables $X = \{x_{i}|1 \leq i \leq m\}$ at the beginning, and $x_{i} = 1$ means that a hybrid edge server is co-located with the AP $\upsilon_{i}$; otherwise $x_{i} = 0$. If $x_{j'} = 1$, we let $y_{i',j'} \geq 0$ indicates the number of service requests from AP $\upsilon_{i'}$ that are routed to the hybrid edge server at $\upsilon_{j'}$; otherwise $y_{i',j'} = 0$.

To reach our goal, we will naturally formulate our problem as an \emph{mixed integer non-linear program (MINLP)}, called the \emph{Optimal Request Scheduling over Hybrid Edge Server Placement (ORS-HESP)} problem:
\begin{subequations}\label{equ-1}
\begin{align}
\label{f1} & \textbf{min} && \sum_{i=1}^{n}\lambda\zeta_{i}+\sum_{i=1}^{n}\pi_{i}\chi_{i}+\sum_{i=1}^{n}\sum_{j=1}^{n}\xi_{i,j}y_{i,j}\\
\label{f2} & \textbf{s.t.} && \sum_{i=1}^{n}x_{i}=m, x_{i}\in\{0,1\}\\
\label{f3} &&& \lfloor\beta\theta_{i}\rfloor\leq\lfloor\alpha K\rfloor,   \forall x_{i}=1\\
\label{f4} &&& \sum_{j=1}^{n}y_{j,i}-\zeta_{i}\leq\lfloor(1-\alpha) K\rfloor, \forall x_{i}=1\\
\label{f5} &&& \sum_{j=1}^{n}y_{i,j}+\lfloor\beta\theta_{i}\rfloor\leq W,  \forall x_{i}=1 \\
\label{f6} &&& \sum_{j=1}^{n}y_{i,j}\leq W,   \forall x_{i}=0\\
\label{f10} &&& \sum_{x_{j}=1}y_{i,j}=\lfloor\chi_{i}-\beta\theta_{i}\rfloor, \forall x_{i}=1\\
\label{f11} &&& \sum_{x_{j}=1}y_{i,j}=\lfloor\chi_{i}\rfloor, \forall x_{i}=0\\
\label{f7} &&& \zeta_{i}=\max\{0,\sum_{j=1}^{n}y_{j,i}-\lfloor(1-\alpha)K\rfloor\}\\
\label{f8} &&& \chi_{i}=\min\{\theta_{i},W\}\\
\label{f9} &&& i,j \in \{1,2,\cdots,n\}
\end{align}
\end{subequations}

\vspace{-0.05in}
The objective (\ref{f1}) minimizes the total delay of the communication network. Constraint (\ref{f9}) specifies the range of values of the parameters $i$ and $j$. Constraint (\ref{f2}) ensures that there are only $m$ hybrid edge servers at APs. Constraints (\ref{f3}) and (\ref{f4}) specify the upper bounds of private computation capacity and public computation capacity, respectively. Constraint (\ref{f5}) and (\ref{f6}) prescribe that the communication capacity should be met at any AP no matter whether there is a co-located hybrid edge server or not. Constraint (\ref{f10}) and (\ref{f11}) declare that the number of user requests that can be served at each edge server equals to the sum of the number of requests dispatched from this AP to other APs which co-locate with hybrid edge servers. Finally, Constraint (\ref{f7}) and (\ref{f8}) present the mathematically definition of $\zeta_{i}$ and $\chi_{i}$ respectively.

\subsection{Complexity Analysis}

We prove that the ORS-HESP problem is NP-hard with a reduction process from the well-known \emph{the Exact Cover Problem}~\cite{Chang2003Solving}, which is a typical NP-complete problem.

\begin{definition}[Exact Cover]
The Exact Cover Problem is described as that given a collection $S$ of subsets of a set $X$, an exact cover of $X$ is a subcollection $S^{*}$ of $S$ that satisfies the following two conditions:
\begin{itemize}
  \item The intersection of any two distinct subsets in $S^{*}$ is empty, i.e., the subsets in $S^{*}$ are pairwise disjoint. In other words, each element in $X$ is contained in at most one subset in$S^{*}$.
  \item The union of the subsets in $S^{*}$ is $X$, i.e., the subsets in $S^{*}$ cover $X$. In other words, each element in $X$ is contained in at least one subset in $S^{*}$.
\end{itemize}
\end{definition}

\begin{theorem}\label{theorem1}
Our ORS-HESP problem in WMAN is NP-hard.
\end{theorem}

\begin{IEEEproof}\label{proof1}
The Exact Cover Problem is a decision problem to determine if an exact cover exists. For the sake of convenience, we first simplify the original problem and consider a special case. We assume that the communication resources at each AP and the total computing resources at edge servers are sufficient. In this case, all the requirements can be met by the collaboration between hybrid edge servers. Therefore, it is not necessary to redirect requests to the data center. As a consequence, all the $\Gamma$ requests are met and processed in $m$ hybrid edge servers through scheduling. We regard the set $X$ of all requests as the Exact Cover Problem elements set, and the number $m$ of APs co-located with hybrid edge servers as the cardinality of the subcollection $S^{*}$. When a request is routed to an exact AP, it means that the element will be included in the exact subset in $S^{*}$. Then the ORS-HESP problem in WMAN is to schedule all $\Gamma$ requests to the $m$ edge servers with minimized delay is equivalent to determine an exact subcollection $S^{*}$, such that each element in $X$ is contained in exactly one subset in $S^{*}$. Thus Theorem \ref{theorem1} is proved.
\end{IEEEproof}

\section{Partition-based Optimization}\label{method}
In this section, we dissolve the complexity of the problem through the analysis of resource competition relationship. By using the \emph{branch and bound} (\textbf{BnB})~\cite{Narendra1977A} approach, our partition-based optimization algorithm is proposed.

\subsection{Analysis of Resource Competition}\label{P-2}

For a series of resource constraints proposed by our model, we must give a comprehensive analysis of the competitive relationship of resources to lay the groundwork for our efficient solution.

\begin{figure}[t]
   \centering
      \includegraphics[width=3.5 in,scale=1.0]{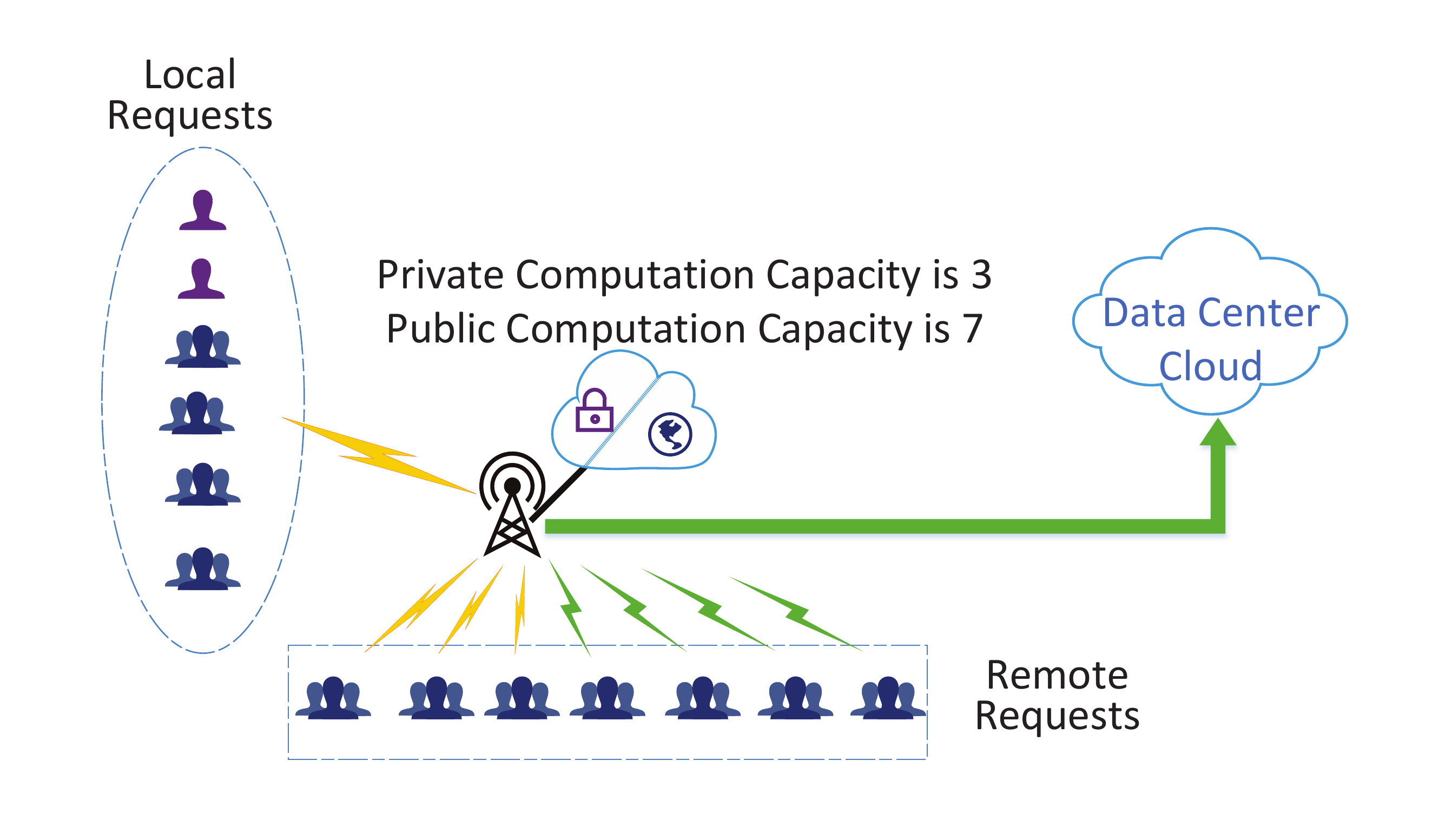}\vspace{-0.1in}
 \caption{The illustrative example of the computation resource competition.}
 \label{fig-competition1}
\end{figure}

\textbf{Competition Rule for Computation Resources:}
We assume that there exists the computation resources competition at the hybrid edge servers. The local requests, which are sent from the locally covered users, will be processed at the hybrid edge servers co-located with local APs. There is no competition between local public requests and private requests since they are processed at different part (public or private). If that edge server still has left resources, the remote public requests from other APs can be routed to this edge server. Since we only consider the queue in a static state, requests will be handled within a time window. For any hybrid edge server, the number of received service requests cannot exceed the computation resource limit, and requests get the computation resource according to the rule of first-come-first-served, i.e., on the basis of the length of the path or the size of the delay. As is shown in Fig. ~\ref{fig-competition1}, we can see that the exceeded requests will be dispatched to the data center cloud since there has unlimited computing power to handle all the requests.

\begin{figure}[t]
   \centering
      \includegraphics[width=3.5 in,scale=1.0]{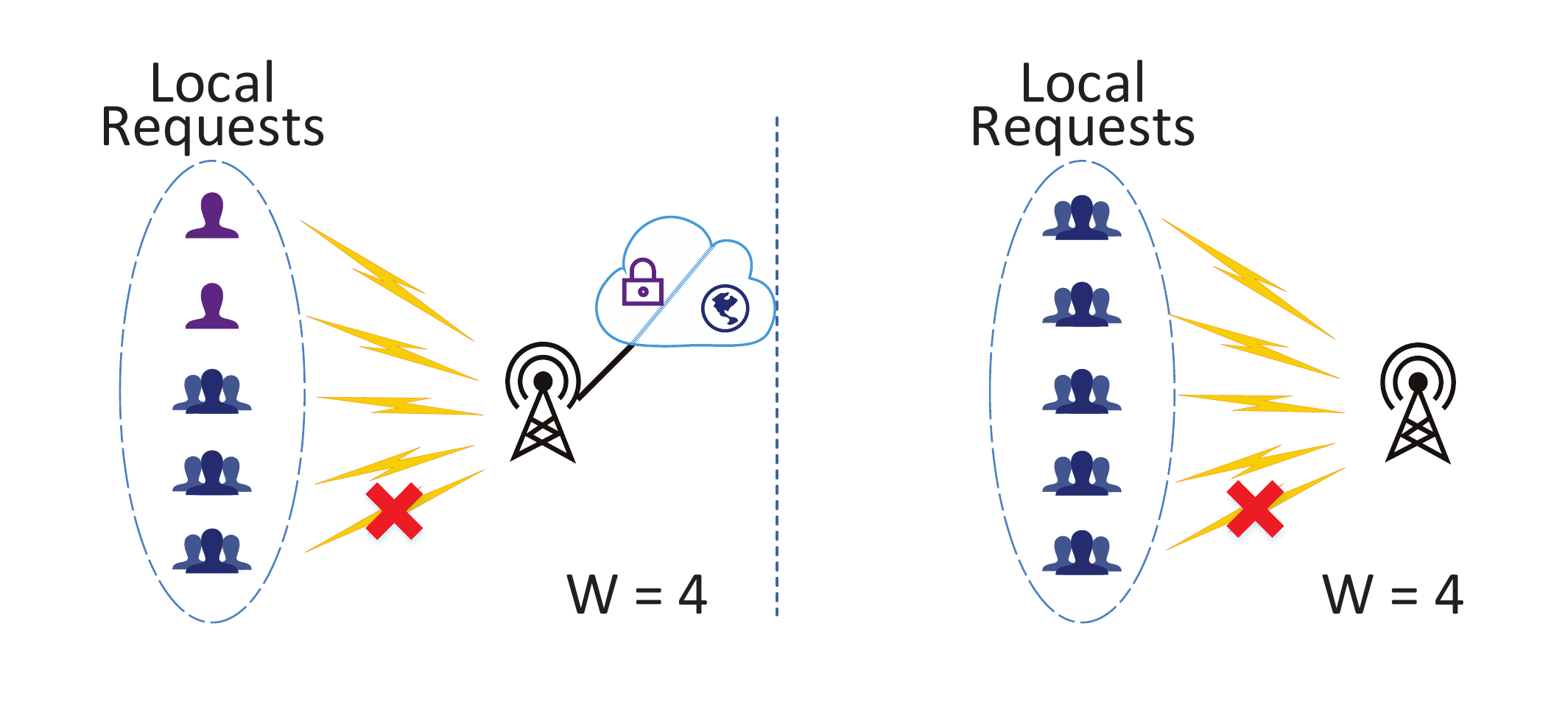}\vspace{-0.1in}
 \caption{The illustrative example of the communication resource competition.}
 \label{fig-competition2}
\end{figure}
\begin{figure}[t]
   \centering
      \includegraphics[width=3.5 in,scale=1.0]{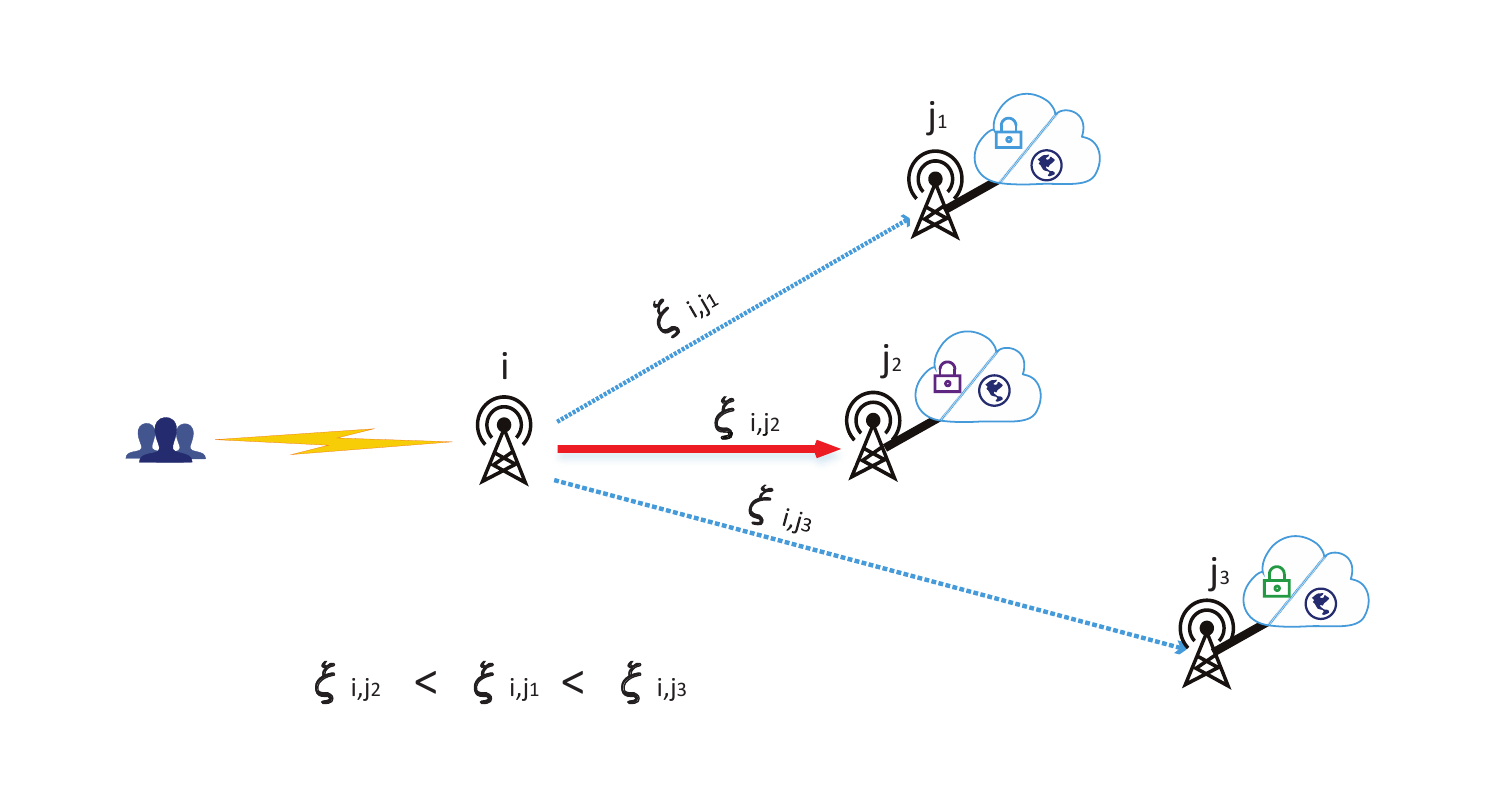}\vspace{-0.1in}
 \caption{The scheduled request will select the path with the least delay.}
 \label{fig-path}
\end{figure}

\textbf{Competition Rule for Communication Resources:}
Now we consider the local communication resource competition due to the limited amount of communication capacity at each AP. Note that all local requests sharing the same bandwidth to send or receive packets. However, due to the security and importance of the private request, the communication of the private request is first satisfied in a time window, and then we strive to satisfy the communication of the public requests. The local public request is treated equally and served as much as possible under the premise of communication capacity. When the number of local public requests exceeds the limit, as Fig. ~\ref{fig-competition2} shows, the extra requests will be blocked to the next time window. There is no competition for communication resources among APs or between the APs and the data center since the bandwidth is by default sufficient to meet the communication demand of all requests.

\textbf{Scheduling Path Selection:}
When the received public request comes from an AP without the edge server or exceeds the local computing capacity, the request should be scheduled to other edge servers with sufficient processing power. As the case diagram illustrates in Fig. ~\ref{fig-path}, we depict $\upsilon_{i}$ as the AP without the edge server and $\upsilon_{j_{1}}$, $\upsilon_{j_{2}}$, $\upsilon_{j_{3}}$ as the AP with the edge server. When the public request from $\upsilon_{i}$ needs to be scheduled to the edge server, we should select the path with the minimum delay since we aim to seek a scheduling strategy for minimizing total delay. In this case, we express the size of the delay by the length of the distance, then we find that $\xi_{i,j_{2}} < \xi_{i,j_{1}} < \xi_{i,j_{3}}$, which means that we will select the path from $\upsilon_{i}$ to $\upsilon_{j_{2}}$ and that the public request will be processed at edge server co-located with $\upsilon_{j_{2}}$.

\subsection{Problem Partitioning}\label{P-3}
When considering the ORS-HESP problem in WMAN, we first go back to the placement of hybrid edge servers. Hybrid edge servers are mainly located nearby a large number of private service requirements, then being located according to the number of public service requirements. Although there are $\binom{n}{m}$ possible combinations of the $m$ hybrid edge servers and the $n$ APs, we still can select a specific arrangement in the light of the popularity of public/private service requirements. We assume that the placement problem is non-deterministic and pre-set for the ORS-HESP problem.

According to the above analysis, the ORS-HESP problem is a quite tricky NP-hard problem. However, we can greatly degrade its complexity by clarifying resource constraints.

\subsubsection{The ORS-HESP Problem with Sufficient Computation Resources and Communication Resources (SKSW-ORS-HESP)}

In this case, computation resources and communication resources are sufficient, so that no request will be dispatched to the data center and all the requests will be processed in the hybrid edge servers, i.e., $\zeta_{i}=0$, $\chi_{i}=\theta_{i}$. There will be no remote delay between the hybrid edge servers and the data center, thus we rewrite the ORS-HESP problem as:

\begin{subequations}\label{equ-2}
\begin{align}
\label{f21}& \textbf{min} && \sum_{i=1}^{n}\pi_{i}\theta_{i}+\sum_{i=1}^{n}\sum_{j=1}^{n}\xi_{i,j}y_{i,j}\\
\label{f22}& \textbf{s.t.} &&(\ref{f2}),(\ref{f3}),(\ref{f9})\\
\label{f24}&&&\sum_{j=1}^{n}y_{j,i}\leq\lfloor(1-\alpha) K\rfloor, \forall x_{i}=1\\
\label{f210}&&&\sum_{x_{j}=1}y_{i,j}=\lfloor(1-\beta)\theta_{i}\rfloor, \forall x_{i}=1\\
\label{f211}&&&\sum_{x_{j}=1}y_{i,j}=\lfloor\theta_{i}\rfloor, \forall x_{i}=0\\
\label{f25}&&&\theta_{i}\leq W,  \forall i \in \{1,2,\cdots,n\}
\end{align}
\end{subequations}
\vspace{-0.1in}
\subsubsection{The ORS-HESP Problem with Insufficient Computation Resources and Sufficient Communication Resources (IKSW-ORS-HESP)}

In this sub-problem, the computation resources are insufficient while the communication resources are sufficient. In this case, there always exists some servers such that $\exists x_{i}=1$ and $\sum_{j=1}^{n}y_{j,i}>\lfloor(1-\alpha) K\rfloor$. On the contrary, all service requests within the coverage of APs will be processed by scheduling to somewhere, hybrid edge servers or data center during the current time window, i.e., $\chi_{i}=\theta_{i}$. Consequently, we retype the ORS-HESP problem as:

\begin{subequations}\label{equ-3}
\begin{align}
\label{f41}& \textbf{min} &&  \sum_{i=1}^{n}\lambda\zeta_{i}+\sum_{i=1}^{n}\pi_{i}\theta_{i}+\sum_{i=1}^{n}\sum_{j=1}^{n}\xi_{i,j}y_{i,j}\\
\label{f42} & \textbf{s.t.} &&(\ref{f2}),(\ref{f3}),(\ref{f9}),(\ref{f7})\\
\label{f44}&&&\sum_{j=1}^{n}y_{j,i}>\lfloor(1-\alpha) K\rfloor,\exists  i \in \{i\mid x_{i}=1, i=1,2,\cdots,n\}\\
\label{f410}&&&\sum_{x_{j}=1}y_{i,j}\leq\lfloor(1-\beta)\theta_{i}\rfloor, \forall x_{i}=1\\
\label{f411}&&&\sum_{x_{j}=1}y_{i,j}\leq\lfloor\theta_{i}\rfloor, \forall x_{i}=0\\
\label{f412}\nonumber&&&\sum_{i=1}^{n}\sum_{x_{j}=1}y_{i,j}+\zeta_{i}=\sum_{i=1}^{n}\lfloor(1-\beta)\theta_{i}\rfloor-\sum_{x_{j}=1}\lfloor(1-\alpha)K\rfloor,\\
&&&\forall i \in \{1,2,\cdots,n\}\\
\label{f45}&&&\theta_{i}\leq W,   \forall i \in \{1,2,\cdots,n\}
\end{align}
\end{subequations}

\vspace{-0.1in}
\subsubsection{The ORS-HESP Problem with Sufficient Computation Resources and Insufficient Communication Resources (SKIW-ORS-HESP)}

In this sub-problem, the computation resources are sufficient but the communication resources are insufficient. This context implies $\zeta_{i}=0$. The computation resource constraint indicates that not every AP can satisfy the communication of its local requests, thus the exceed part will be delayed, i.e., $\exists  i \in \{1,2,\cdots,n\}$, $\theta_{i}> W$. Then we rewrite the ORS-HESP problem as:

\begin{subequations}\label{equ-4}
\begin{align}
\label{f31}& \textbf{min} && \sum_{i=1}^{n}\pi_{i}\chi_{i}+\sum_{i=1}^{n}\sum_{j=1}^{n}\xi_{i,j}y_{i,j}\\
\label{f32} & \textbf{s.t.} &&(\ref{f2}),(\ref{f3}),(\ref{f10}),(\ref{f11}),(\ref{f8}),(\ref{f9})\\
\label{f34}&&&\sum_{j=1}^{n}y_{j,i}\leq\lfloor(1-\alpha) K\rfloor, \forall x_{i}=1\\
\label{f35}&&&\theta_{i}> W,   \exists  i \in \{1,2,\cdots,n\}
\end{align}
\end{subequations}
\vspace{-0.1in}
\subsubsection{The ORS-HESP Problem with Insufficient Computation Resources and Communication Resources (IKIW-ORS-HESP)}

In this case, both the computation resources and communication resources are insufficient. As a result, some requests with insufficient communication resources, under the coverage of APs, will be delayed. Then the number of all public requests still exceeds the total computing capacity, leading to the redirection of the rest requests to the remote data center. Based on the above description, we have constraints (\ref{f44}) and (\ref{f35}). Finally, we rewrite the ORS-HESP problem as:

\begin{subequations}\label{equ-5}
\begin{align}
\label{f51}&&&(\ref{f1})\\
\label{f52} & \textbf{s.t.} &&(\ref{f2}),(\ref{f3}),(\ref{f7}),(\ref{f8}),(\ref{f9}),(\ref{f44}),(\ref{f35})\\
\label{f510}&&&\sum_{x_{j}=1}y_{i,j}\leq\lfloor\chi_{i}-\beta\theta_{i}\rfloor,\forall x_{i}=1\\
\label{f511}&&&\sum_{x_{j}=1}y_{i,j}\leq\lfloor\chi_{i}\rfloor, \forall x_{i}=0\\
\label{f512}\nonumber&&&\sum_{i=1}^{n}\sum_{x_{j}=1}y_{i,j}=\sum_{i=1}^{n}\lfloor(1-\beta)\theta_{i}\rfloor-\sum_{x_{j}=1}\lfloor(1-\alpha)K\rfloor,\\ &&&\forall i \in \{1,2,\cdots,n\}
\end{align}
\end{subequations}

\vspace{-0.1in}

\subsection{Our Algorithm}\label{5-1}
To solve the ORS-HESP problem efficiently, we have divided it into four sub-problems and formulated the corresponding MILP expressions. Given the communication network $G= (V, E)$, the placement of hybrid edge servers will be decided mainly according to the service popularity or the regional particularity. Then we decide whether or not to postpone part of the requests by examining communication resources constraint at each AP. After solving the access problem, we determine whether it is necessary to schedule part of the requests to the data center by checking computation resources constraint of each hybrid edge server. The detailed algorithm pseudocode is given in \textbf{Algorithm ~\ref{algo1}}. In addition to clarifying resource constraints, we can totally transform the model from \textbf{non-linear} to \textbf{linear} by determining the values of $\zeta_{i}$ and $\chi_{i}$. Based on this insight, we use the branch and bound approach to solve the ORS-HESP problem, and Section~\ref{evaluation} proves that the partition-based optimization algorithm has excellent performance.

\begin{algorithm}[t]
\caption{The Partition-based Optimization Algorithm }
\label{algo1}
\small
\begin{algorithmic}[1]
\REQUIRE A communication network $G=(V,E)$.
\ENSURE Optimal scheduling scheme and minimum delay.
\STATE $x_{i}\longleftarrow\{0,1\}$
\STATE $/*$ appoint $m$ APs co-located with hybrid edge servers. $*/$
\FOR  {each AP$i$}
\IF   {$\forall i \in \{1,2,\cdots,n\}$,$\theta_{i}\leq W$}
\IF   {$Pu\leq m\lfloor(1-\alpha) K\rfloor$}
\STATE    $/*$ $Pu$ denotes all public requests left to be processed. $*/$
\STATE    $\zeta_{i}\longleftarrow0$, $\chi_{i}\longleftarrow\theta_{i}$
\STATE    solve \textbf{Model} $(\ref{equ-2})$.
\ELSE
\IF   {$\sum_{j=1}^{n}y_{j,i}>\lfloor(1-\alpha) K\rfloor$}
\STATE    $\zeta_{i}\longleftarrow\sum_{j=1}^{n}y_{j,i}-\lfloor(1-\alpha) K\rfloor$
\ELSE
\STATE    $\zeta_{i}\longleftarrow0$
\ENDIF
\STATE    solve \textbf{Model} $(\ref{equ-3})$.
\ENDIF
\ELSE
\IF   {$Pu\leq m\lfloor(1-\alpha) K\rfloor$}
\IF   {$\theta_{i}\leq W$}
\STATE    $\chi_{i}\longleftarrow\theta_{i}$
\ELSE
\STATE    $\chi_{i}\longleftarrow W$
\ENDIF
\STATE    solve \textbf{Model} $(\ref{equ-4})$.
\ELSE
\IF   {$\theta_{i}\leq W$}
\IF   {$\sum_{j=1}^{n}y_{j,i}>\lfloor(1-\alpha) K\rfloor$}
\STATE    $\chi_{i}\longleftarrow\theta_{i}$, $\zeta_{i}\longleftarrow\sum_{j=1}^{n}y_{j,i}-\lfloor(1-\alpha) K\rfloor$
\ELSE
\STATE    $\chi_{i}\longleftarrow\theta_{i}$, $\zeta_{i}\longleftarrow0$
\ENDIF
\ELSE
\IF   {$\sum_{j=1}^{n}y_{j,i}>\lfloor(1-\alpha) K\rfloor$}
\STATE    $\chi_{i}\longleftarrow W$, $\zeta_{i}\longleftarrow\sum_{j=1}^{n}y_{j,i}-\lfloor(1-\alpha) K\rfloor$
\ELSE
\STATE    $\chi_{i}\longleftarrow W$, $\zeta_{i}\longleftarrow0$
\ENDIF
\ENDIF
\STATE    solve \textbf{Model} $(\ref{equ-5})$.
\ENDIF
\ENDIF
\ENDFOR
\end{algorithmic}
\end{algorithm}

\section{Performance Evaluation}\label{evaluation}
In this section, we evaluate the performance of our algorithm in terms of the time-consumption and number of iterations to seek a proper solution for the ORS-HESP problem under our experiment settings.

\subsection{Settings of Evaluations}

\subsubsection{Datasets}

\begin{figure}[t]
   \centering
      \includegraphics[width=0.7\linewidth]{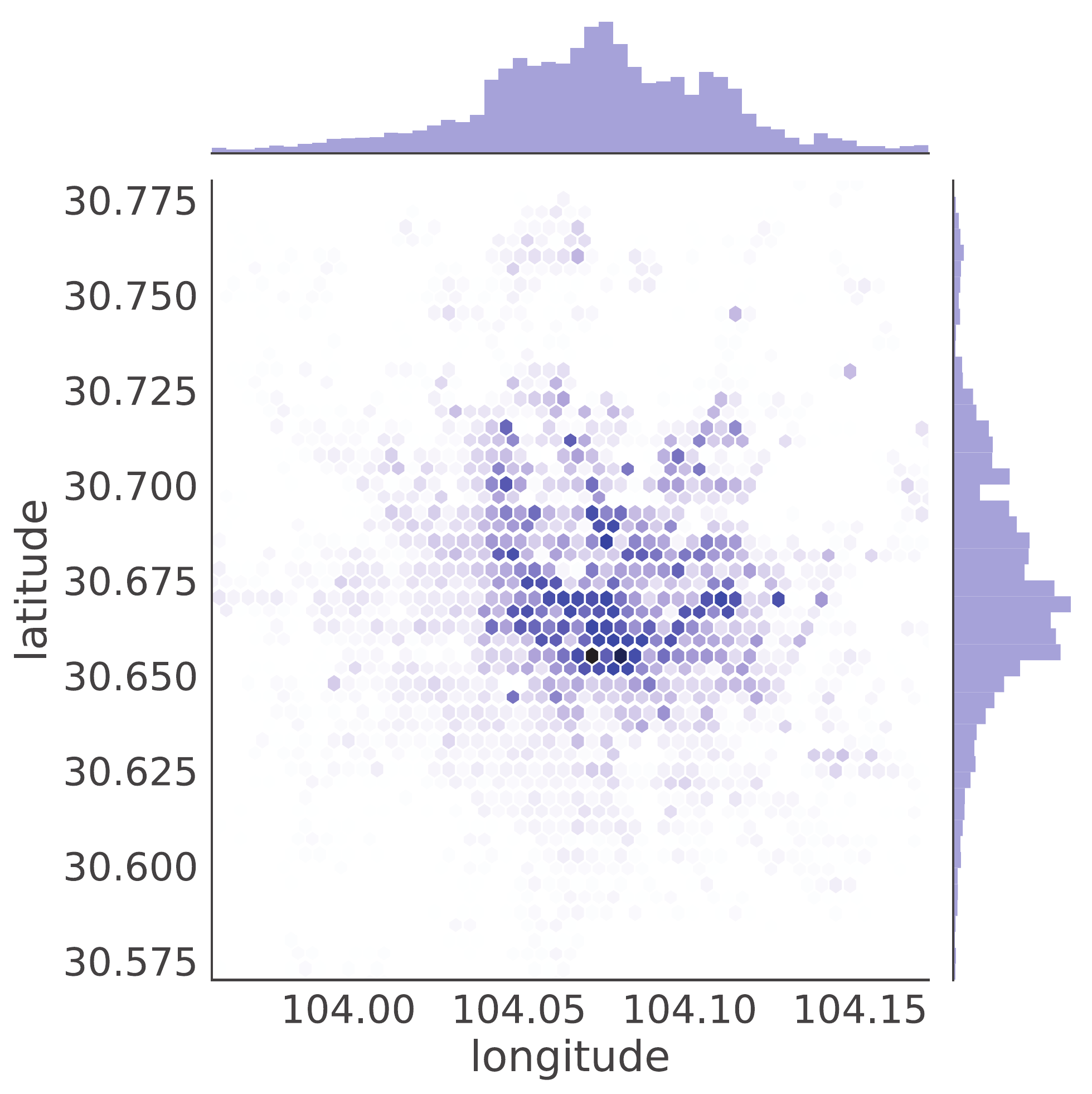}\vspace{-0.1in}
 \caption{The city order request data map.}\vspace{-0.2in}
 \label{distribution}
\end{figure}

Especially, we use the order data sets for Chengdu taxi users, obtained from Didi Chuxing GAIA Initiative ~\cite{Didi}. To be fair enough, we sample one day's data every five days from July 2017 to May 2018 and divide the metropolitan area according to latitude and longitude. Then, we perform experiments over the city order request map as shown in Fig.~\ref{distribution}.

As shown in Fig.~\ref{distribution}, we describe the distribution of user orders through thermodynamics. Obviously, the majority of the sampled orders happened in the middle of the city, because the central location is also the commercial and financial center, and the traffic is large. The total amounts of orders that located in a specific latitude and longitude are also depicted by the bar diagrams on horizontal and vertical axes, respectively.

All the data processing and algorithms are implemented with Python 3.6, and all models are solved by IBM ILOG CPLEX 12.6.1. The experiment environment is configured on a host with Mac OS 64 bits, 2.8GHz Intel Core i7 Processor, and 16 GB 2400 MHz DDR4 memory.

\newcommand{\tabincell}[2]{\begin{tabular}{@{}#1@{}}#2\end{tabular}}
\begin{table*}[t]
\centering
 \caption{PARAMETER SETTINGS}
\begin{tabular}{|c|c|c|c|c|c|c|}
 \hline
   & Parameter &  \multicolumn{5}{c|}{Setting} \\
   \hline
   \hline
 \multirow{7}{*}{\tabincell{c}{Communication \\Network\\ in WMAN}}&Latitude range of the given map & \multicolumn{5}{c|}{[30.57,30.78]}\\
 &Longitude range of the given map & \multicolumn{5}{c|}{[103.96,104.17]}\\
 & $\alpha$ &  \multicolumn{5}{c|}{0.3 in Fig.~\ref{t-size}/Fig.~\ref{t-k}, [0,1] in Fig.~\ref{t-pav}}\\
 & $\beta$ &  \multicolumn{5}{c|}{0.1}\\
 &Local communication delay &  \multicolumn{5}{c|}{Average distance between local users and AP}\\
 &The communication delay between APs &  \multicolumn{5}{c|}{80\% to 120\% of the average distance between APs}\\
 &The remote communication delay &  \multicolumn{5}{c|}{10 times the maximum distance between APs}\\
 \hline
 \hline
 \multirow{3}{*}{\tabincell{c}{Network \\ Size}}& Divided size & 12*10 & 24*20 & 36*30 & 48*40 & 60*50\\
 & The number of APs & 120 & 480 & 1080 & 1920 & 3000\\
 & The number of hybrid edge servers & 40 & 60 & 80 & 100 & 120\\
 \hline
 \end{tabular}
  \label{parametersetting}
 \end{table*}

\subsubsection{Settings of the hybrid edge computing environment}
We take the following three steps to deploy the APs, the hybrid edge servers and set up experimental data.

\textbf{Grid division.} Given a city order request data map within a certain latitude and longitude range, we divide the geographical area into grids with different granularities (s.t., $12*10$, $24*20$, $36*30$, $48*40$, $60*50$). Specifically,  a $12*10$ division means that the latitude range is cut into 12 segments while the longitude range is cut into 10 segments thus finally we get 120 equally sized grids.

\textbf{Location selection.} We select the center coordinate of each grid as the location of the each AP and choose the location of the hybrid edge server based on the number of requests contained within each grid, which means placing hybrid edge servers respectively in the grids with the highest number of requests, since APs and edge servers are often deployed in densely populated areas in the real world.

\textbf{Data preprocessing.} Based on the coordinates of the user orders, we regard each order as a service request with the associated location. In this manner, we generate the requests, APs, and hybrid edge servers in the metropolitan area grid based on the dataset. Thereafter, we treat the average distance of user requests arriving at the local AP in each grid as the communication delay between a user and the AP. Similarly, the distance between any pair of APs is regarded as the inter-AP communication delay. The delay between any AP to the remote data center is fixed as ten times of the maximum inter-AP delay. Without the loss of generality, we finally calculate the average number of orders generated per grid per day as the number of local requests per AP, since a quite large number of service requests are generated even in a small period of time considering the scenario we envision.

Through the above experimental setup, we finally show the parameter data used in the experiment in the TABLE ~\ref{parametersetting}.

\subsection{Methodologies of Experiments}

Following the ``divide and conquer'' philosophy, we divide the original ORS-HESP problem into four sub-problems to ease its complexity. These sub-problems are thereafter solved using the  \emph{the branch and bound (BnB)} theory. In our evaluation, we conduct our experiments from the following aspects to quantify the resulted time-consumption and iteration amount.

\begin{figure*}[t]
\centering
\subfigure[SKSW sub-problem]
{\includegraphics[width=.24\linewidth]{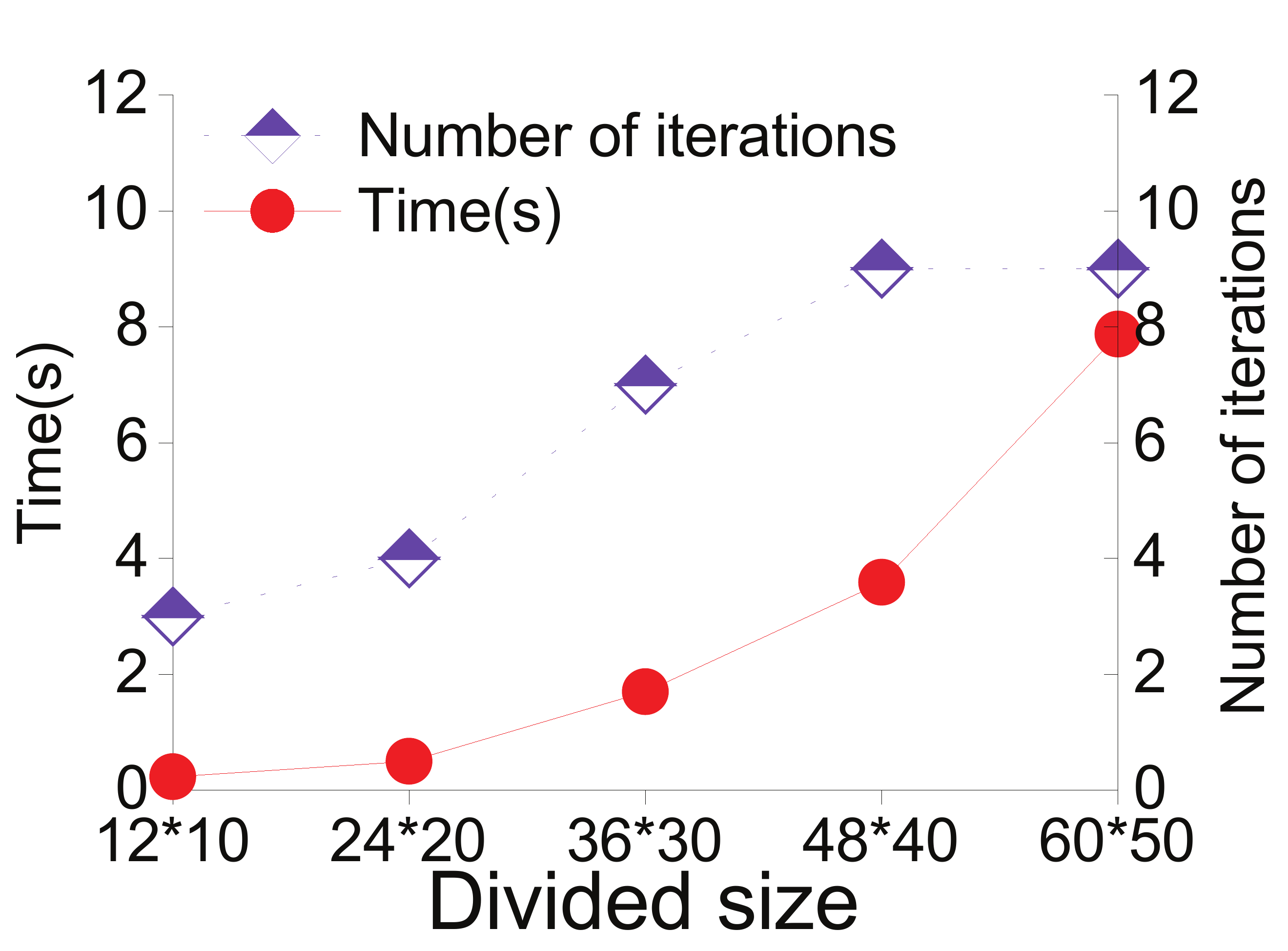}\label{fig:2-1}}
\subfigure[SKIW sub-problem]
{\includegraphics[width=.24\linewidth]{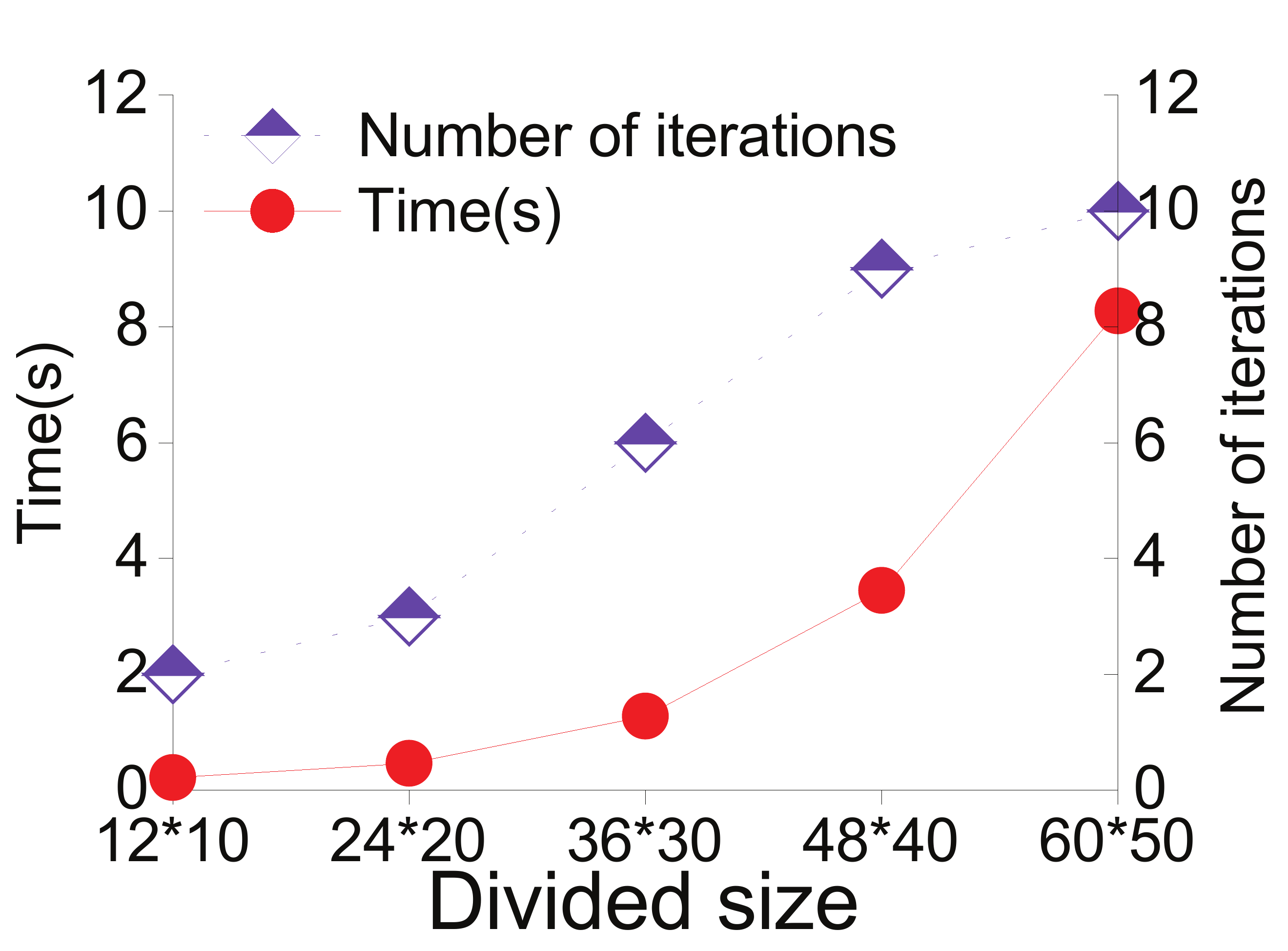}\label{fig:2-2}}
\subfigure[IKSW sub-problem]
{\includegraphics[width=.24\linewidth]{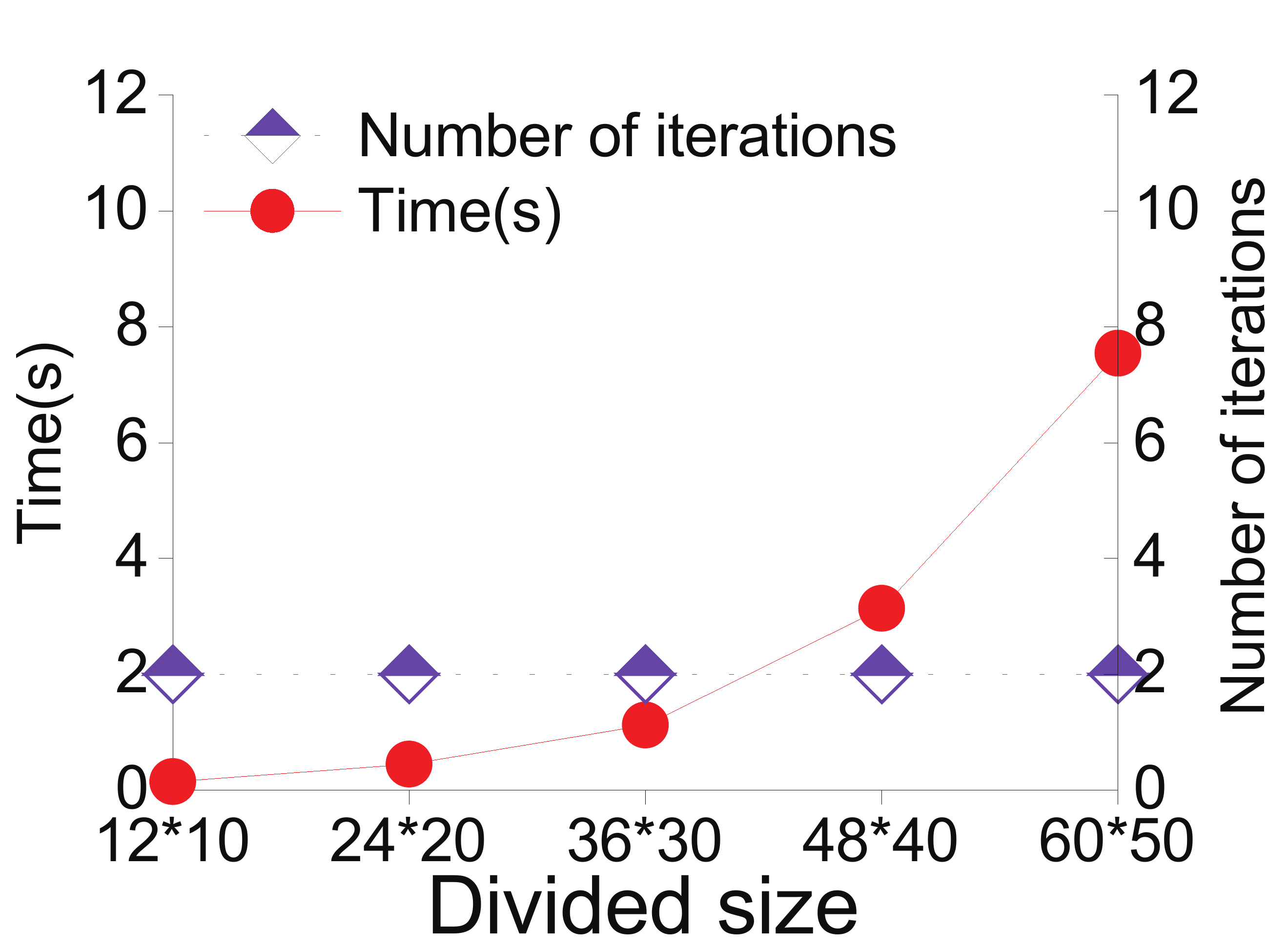}\label{fig:2-2}}
\subfigure[IKIW sub-problem]
{\includegraphics[width=.24\linewidth]{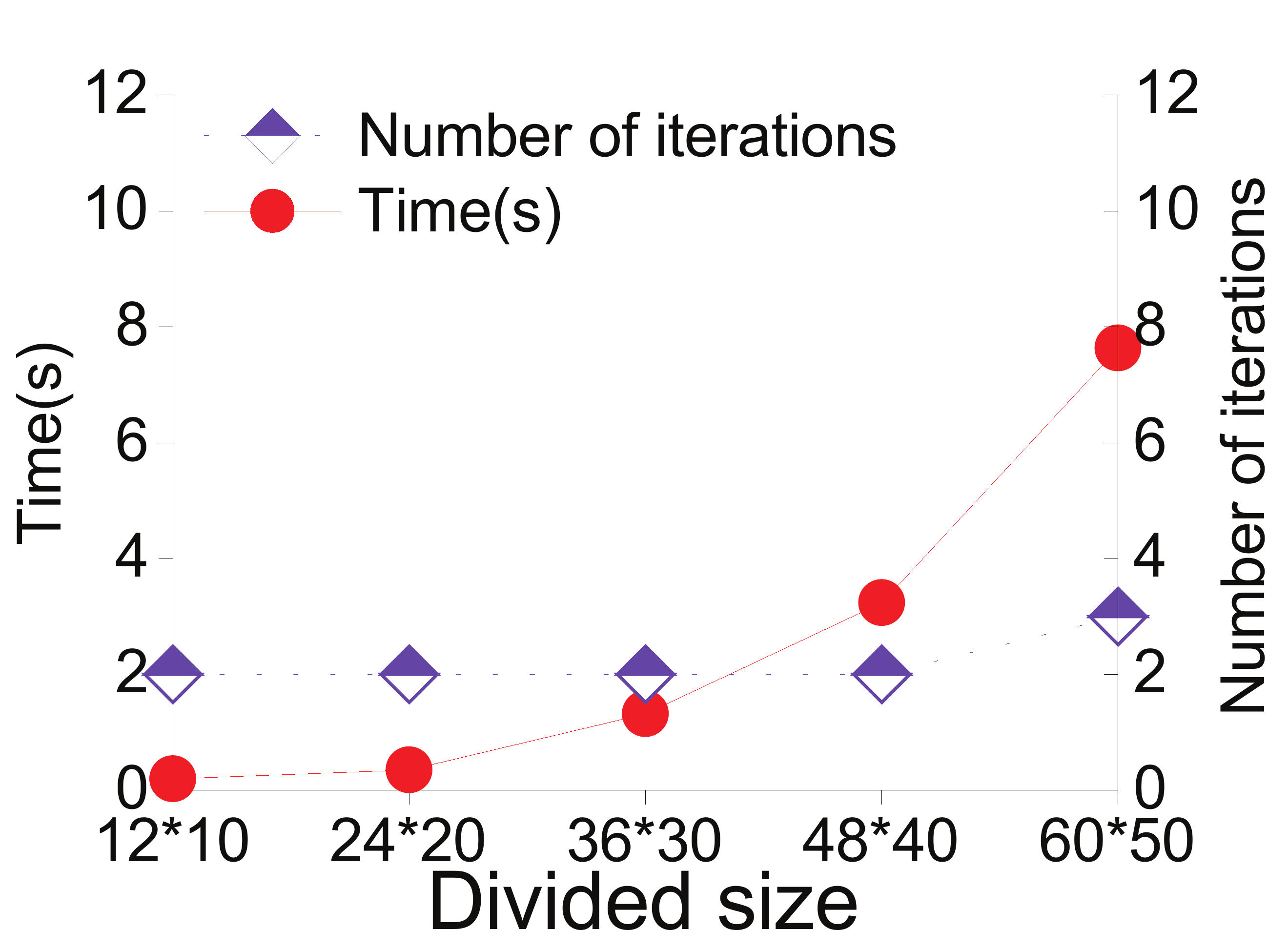}\label{fig:2-2}}
\caption{The effect of grid size on computation time and number of iterations.}
\label{t-size}
\end{figure*}

\textbf{The effect of grid size on the computation time and the number of iterations.} As shown in Fig.~\ref{t-size}, we test the situation of four sub-problems under different divided sizes, obtain the optimal solution and record the number of iterations and the computation time to reach the optimal solution.

\begin{figure*}[t]
\centering
\subfigure[SKSW sub-problem]
{\includegraphics[width=.24\linewidth]{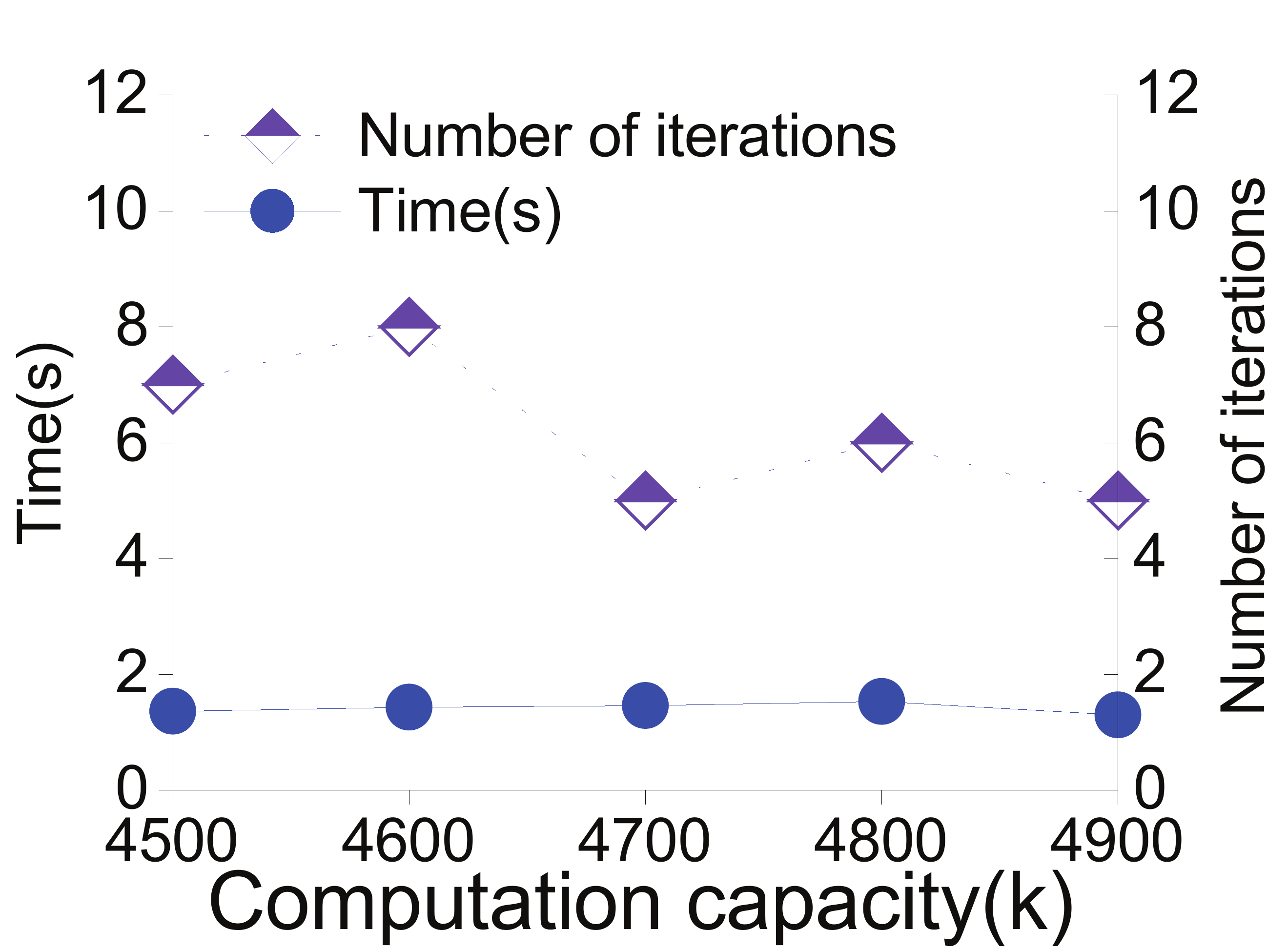}\label{fig:2-1}}
\subfigure[SKIW sub-problem]
{\includegraphics[width=.24\linewidth]{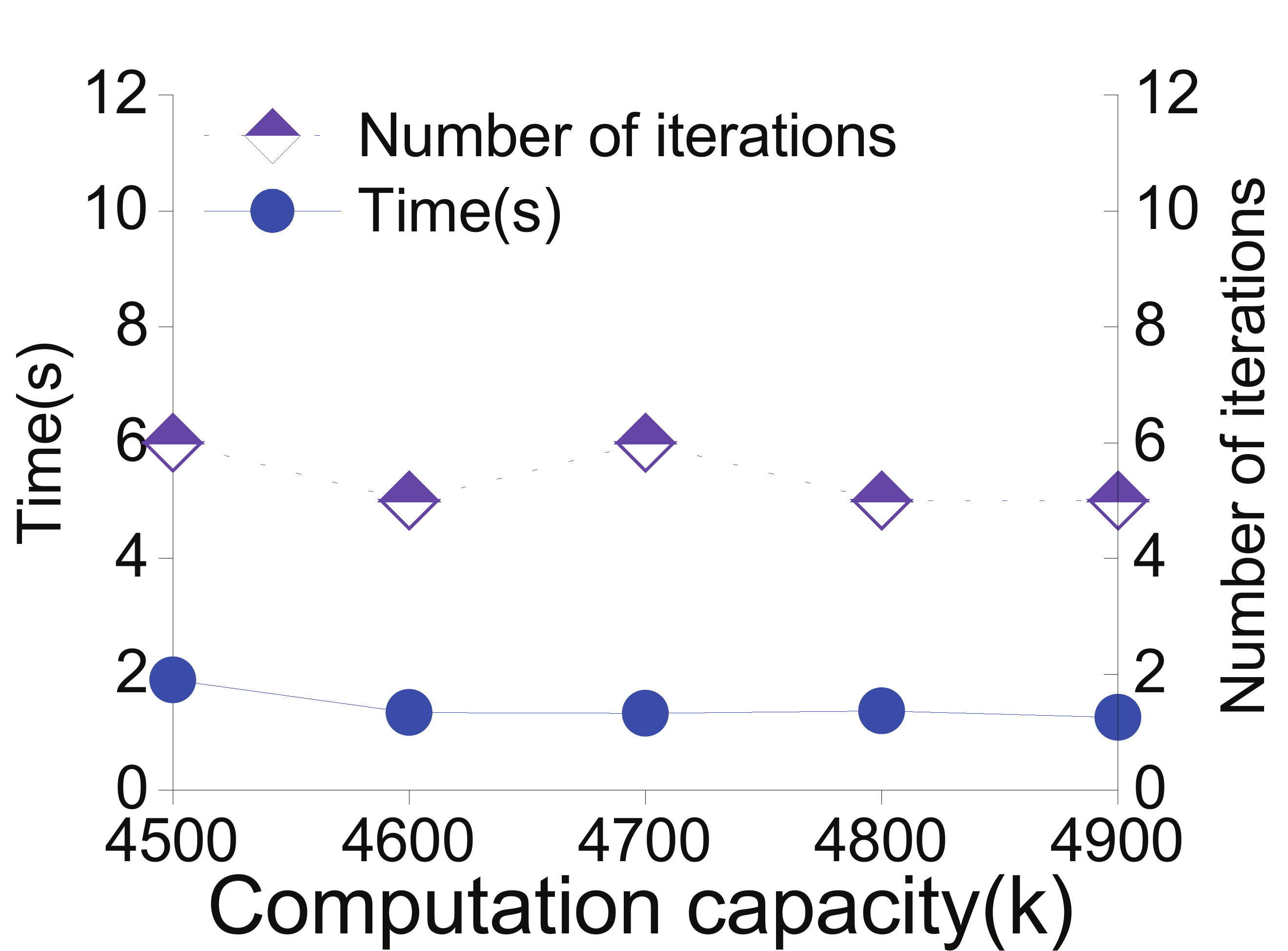}\label{fig:2-2}}
\subfigure[IKSW sub-problem]
{\includegraphics[width=.24\linewidth]{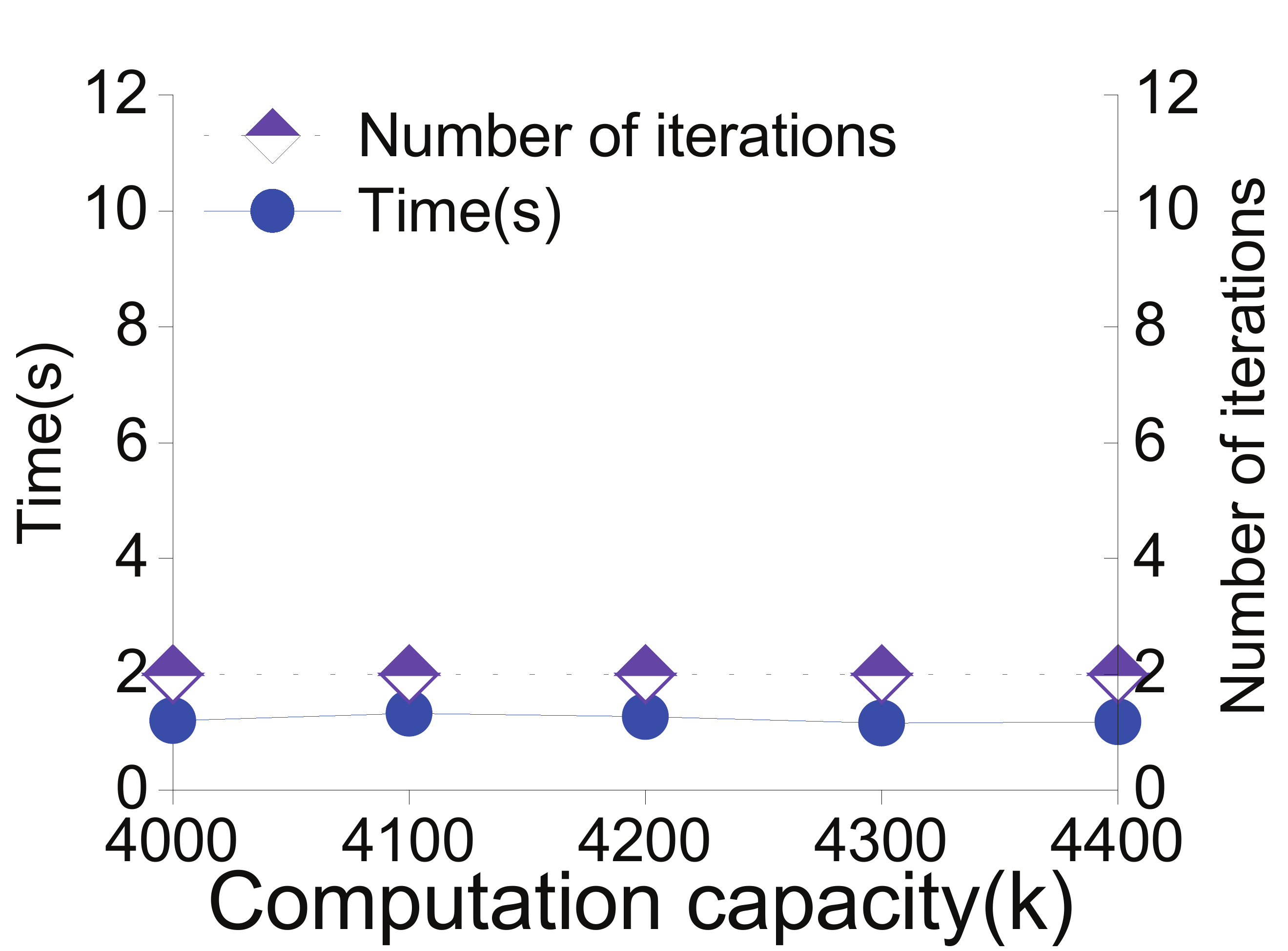}\label{fig:2-2}}
\subfigure[IKIW sub-problem]
{\includegraphics[width=.24\linewidth]{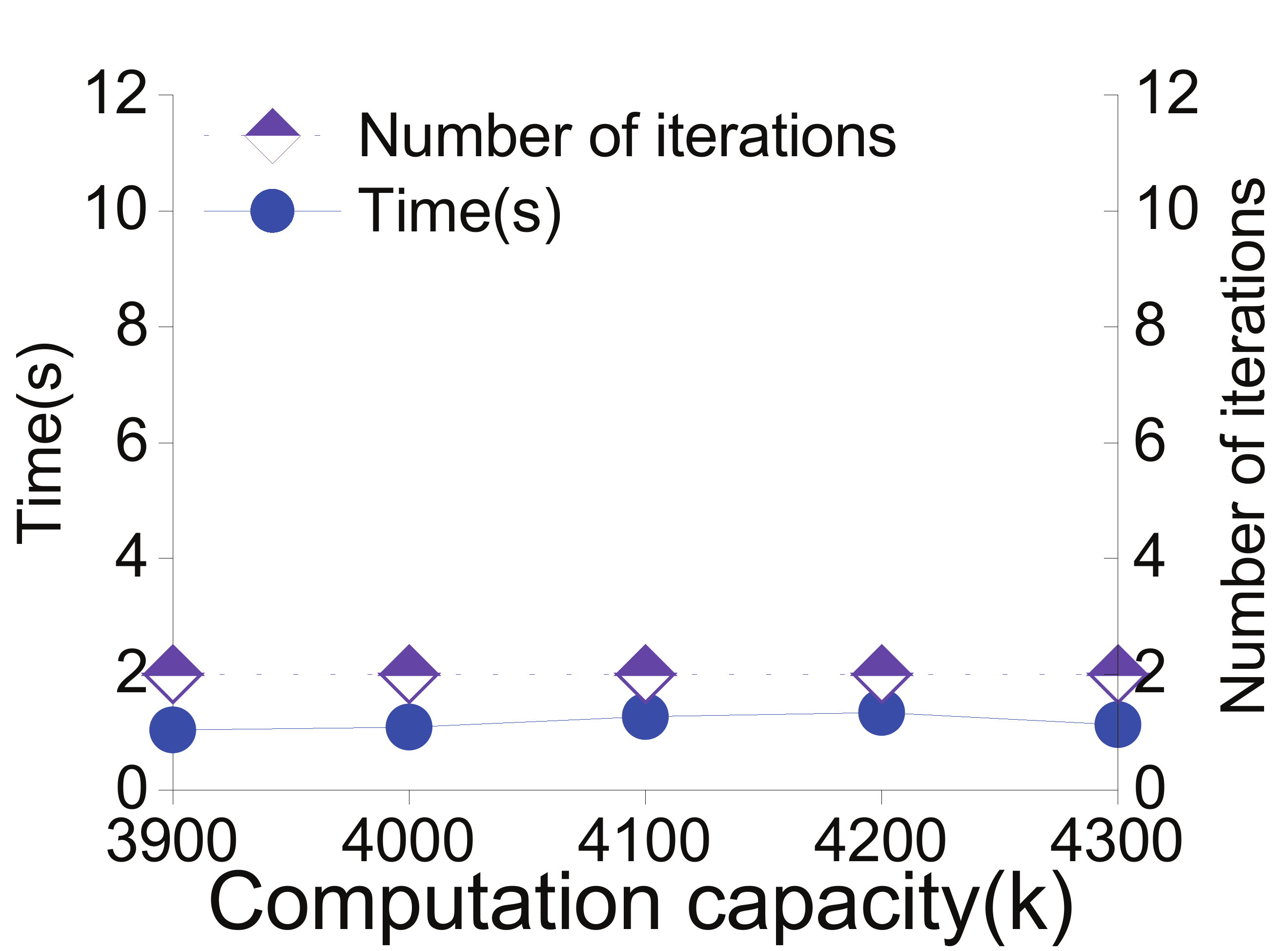}\label{fig:2-2}}
\caption{The effect of computation capacity (k) on computation time and number of iterations.}
\label{t-k}
\end{figure*}

\textbf{The effect of computation capacity ($K$) on the computation time and the number of iterations.} As shown in Fig.~\ref{t-k}, we test the situation of four sub-problems under different values of computation capacity ($K$), obtain the optimal solution and record the number of iterations and the computation time to reach the optimal solution.

\textbf{The effect of $\alpha$ on the the private/public service rate and the average computation time.} As shown in Fig.~\ref{t-pav}, the private/public service rate tells that if all private/public edge service requests can be responded and the average computation time considers all situations with different grid sizes and computation capacities. We test the private/public service rate and the average computation time to show the special property of $\alpha$.

For each situation, we perform five calculations to get the final mean value, and we present the final results in Fig.~\ref{t-size}, Fig.~\ref{t-k}, and Fig.~\ref{t-pav}.

\vspace{-0.1in}
\subsection{Performance Analysis}

Note that in our experiments, we calculate the mean value of 5 executions as the final result. As depicted in Fig.~\ref{t-size}, we study how can the computation time and the number of iterations, which are required to calculate the optimal solution, be affected by the divided size for different sub-problems. In this group of experiments, we determine the value of computation capacity ($K$), according to the total number of user requests and hybrid edge servers. Also, we control the experiment by the communication capacity.

In Fig.~\ref{t-k}, we set the divided size as $36*30$, and study how can the computation time and the number of iterations be affected by the computation capacity ($K$) for different sub-problems. We record the results in line with the adjustment of the value of $K$. For each sub-problem, it is necessary to compute the upper and lower bounds of the value in advance, based on the constraints of each sub-problem and the total number of user requests and servers.

\begin{figure}[t]
\centering
\subfigure[Private/public service rate]
{\includegraphics[width=.49\linewidth]{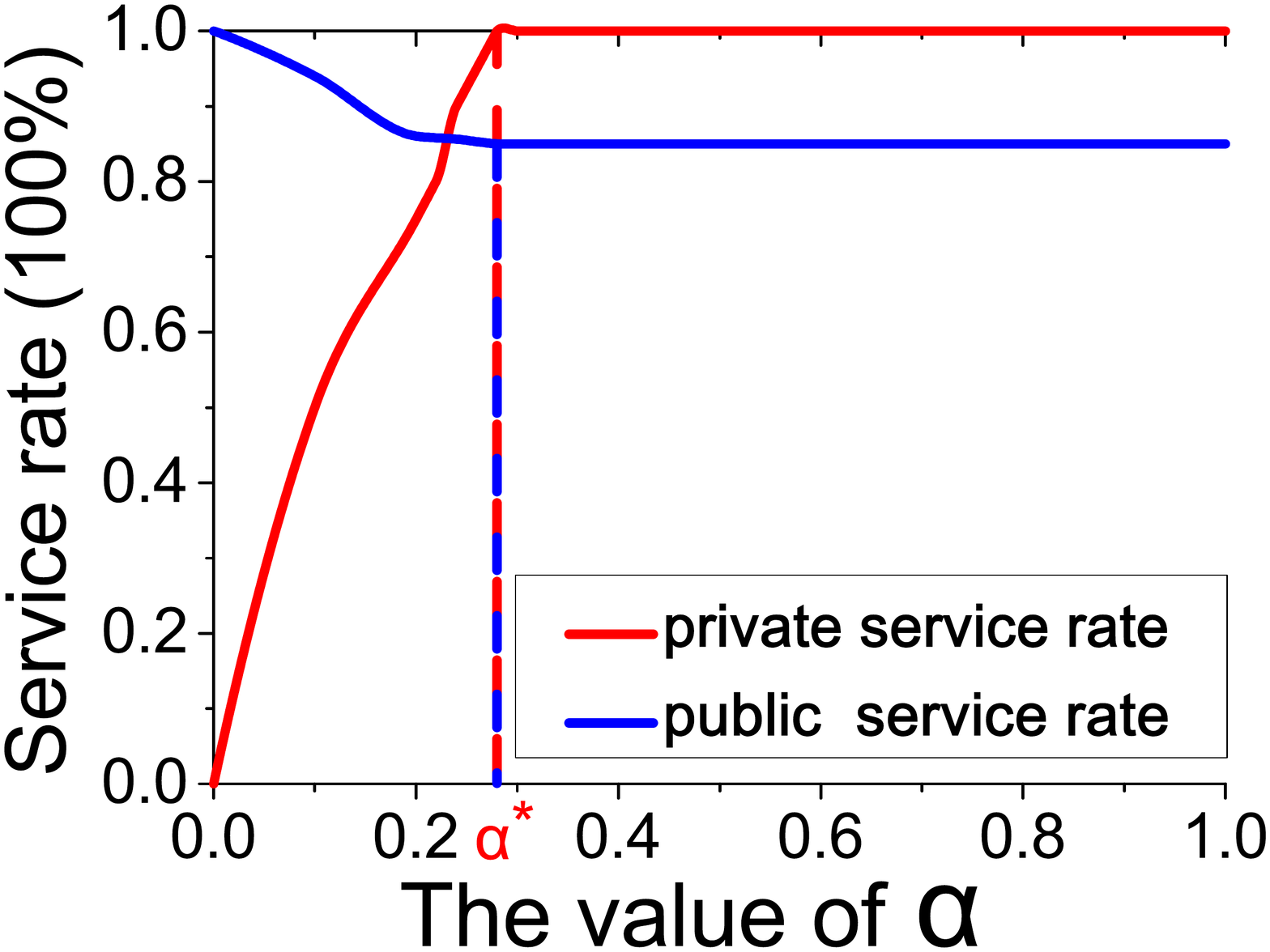}\label{fig:3-1}}
\subfigure[The average computation time]
{\includegraphics[width=.49\linewidth]{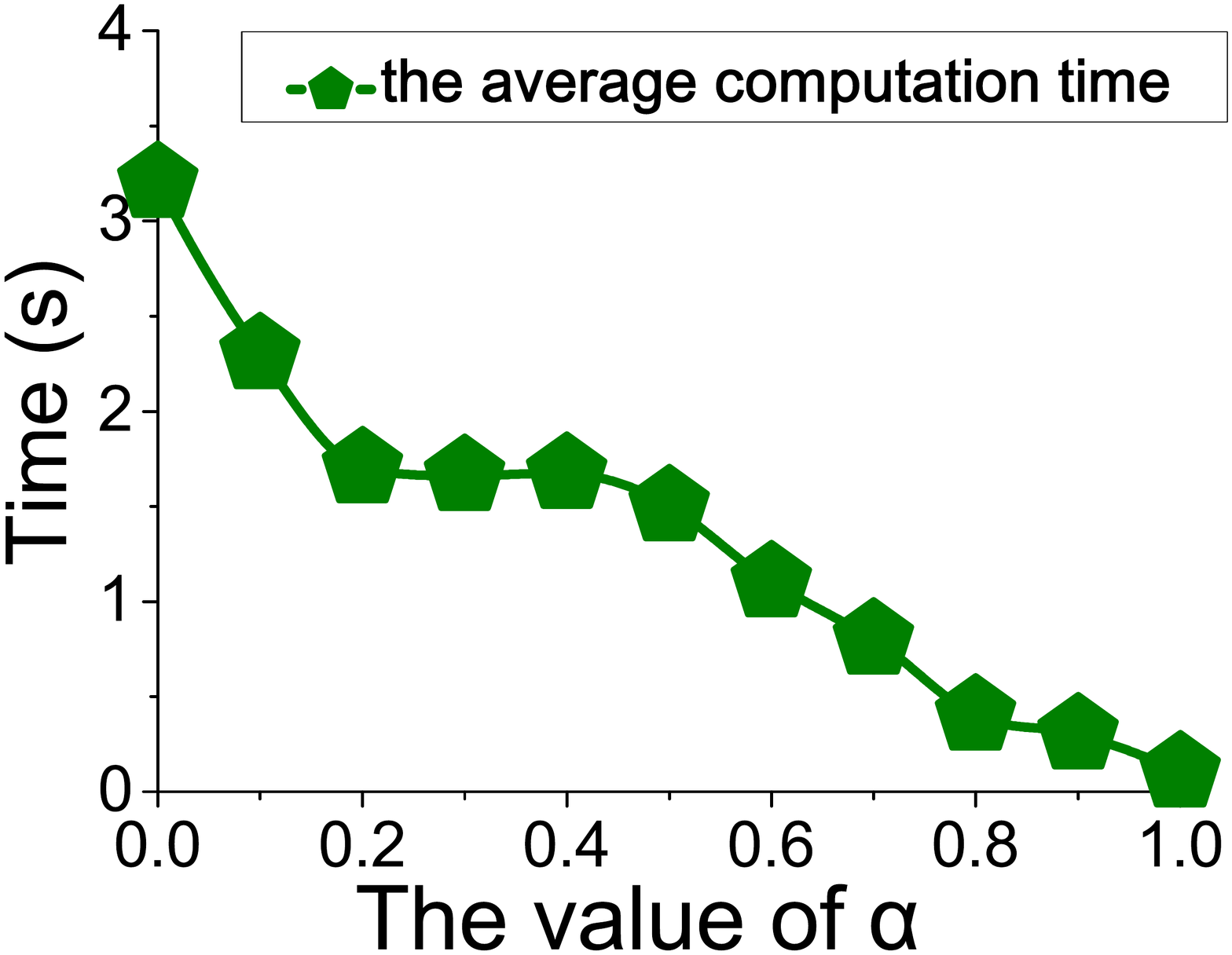}\label{fig:3-2}}
\caption{The effect of \textbf{$\alpha$} on the the private/public service rate and the average computation time.}
\label{t-pav}
\end{figure}

In Fig.~\ref{t-pav}, we consider all situations with different divided sizes and computation capacities for different sub-problems and calculate the mean values to show the effect of $\alpha$. For each situation, we record the number of responded private/public requests to compare with the total requests and the average computation time.
%\begin{figure}[t]
%   \centering
%      \includegraphics[width=3.5 in,scale=1.0]{figure//t-p}\vspace{-0.1in}
%\caption{The effect of \textbf{$\alpha$} on the the private/public service rate and the average computation time.}
%\label{t-p}
%\end{figure}

From the observation of Fig.~\ref{t-size}, Fig.~\ref{t-k} and Fig.~\ref{t-pav}, we can reveal the following facts:
\begin{itemize}
  \item We can conclude that the computation time increases with data size according to the trend of the solid line in Fig.~\ref{t-size} and that the value of computation capacity ($K$) has limited impact on the computation time, according to the trend of the solid line in Fig.~\ref{t-k}.
  \item Regardless of the variations of the data size and computation capacity, the optimal solution for the sub-problem models can be solved quickly. In fact, we spent no more than 9 seconds in the experiments of Fig.~\ref{t-size}, and 2 seconds in Fig.~\ref{t-k}.
  \item In these two sets of comparative experiments, our algorithm obtained the optimal solution with up to 10 iterations, which means that the BnB method is effective to search out the optimal solution.
  \item Fig.~\ref{fig:3-1} shows that when $\alpha \mathrm{<}\alpha^{*}$, the increase of private service rate leads to a significant drop in the public service rate. The reason is that they have to compete for the communication resources. By contrast, when the $\alpha$ exceeds $\alpha^{*}$, then the communication resources for the public requests will be immutable, which leads to the invariability of the public service rate. As depicted in Fig.~\ref{fig:3-2}, the average computation time decreases gradually with the increase of $\alpha$, since the number of requests scheduled between APs goes down.
\end{itemize}

%1) We can conclude that the computation time increases with data size according to the trend of the solid line in four subgraphs of Fig.~\ref{t-size}, and that the value of computation capacity ($K$) has little effect on the computation time, according to the trend of the solid line in four subgraphs of Fig.~\ref{t-k}, since there is no obvious trend of change from the observation.
%
%2) Regardless of the data size and how the computation capacity of the hybrid edge server changes, we still can find the optimal solution for the sub-problem models in a very short time. In fact, we spent no more than 9 seconds at most in the twenty experiments of Fig.~\ref{t-size}, and 2 seconds in the case of Fig.~\ref{t-k}.
%
%3) In these two sets of comparative experiments, our algorithm obtained the optimal solution by performing up to 10 iterations, which means that \emph{the branch and bound method} is very fast and effective for searching the optimal solution of the sub-problem model.
%
%4) By splitting the original ORS-HESP problem into four sub-problems and using the \emph{the branch and bound method}, we eventually achieve the optimal solution with much less complexity.

As a summary, our algorithm is competent to derive out the optimal solution of the ORS-HESP problem during a short time with several rounds of iterations.
Since the computation time is still very short for large-scale operations, we prove that our algorithm can solve the ORS-HESP problem very effectively.
The experimental results show that the algorithm can control the number of iterations to a certain extent and greatly reduce the complexity of the problem. It also proves that the algorithm has very good scalability.

\section{Conclusion and Future Work}\label{conclusionfuturework}

In this work, we put forward the framework of hybrid edge computing. Then we raise the Optimal Request Scheduling over Hybrid Edge Server Placement (ORS-HESP) problem in the hybrid edge computing environment. To solve the ORS-HESP problem effectively, we propose our partition-based optimization algorithm to completely split it into four sub-problems and convert the sub-problems from MINLP to MILP. In the performance evaluation, we experiment over a hybrid edge server placement scenario according to realistic considerations, solve the optimal request scheduling problem, and evaluate the effectiveness and scalability of our algorithm.

This paper is the first attempt to study the request scheduling of edge services under the hybrid edge computing environment, and provides a benchmark idea and method to solve this problem. Nevertheless, the problem can be further studied in other directions. For example, i) the service types mentioned in our paper are specified, while in the future work a joint optimization problem involving more types of services may need to be considered; ii) the method proposed in this paper mainly studies the case where the hybrid edge servers are homogeneous, while more complex networks with heterogeneous hybrid edge servers can be further studied in our future work.

\section*{Acknowledgement}
This work is partially supported by National Program for Support of Top-Notch Young Professionals of National Program for Special Support of Eminent Professionals, and National Natural Science Foundation of China under Grant No.61772544.

\bibliographystyle{IEEEtran}
\bibliography{next}

\begin{IEEEbiography}[{\includegraphics[width=1in,height=1.25in,clip,keepaspectratio]{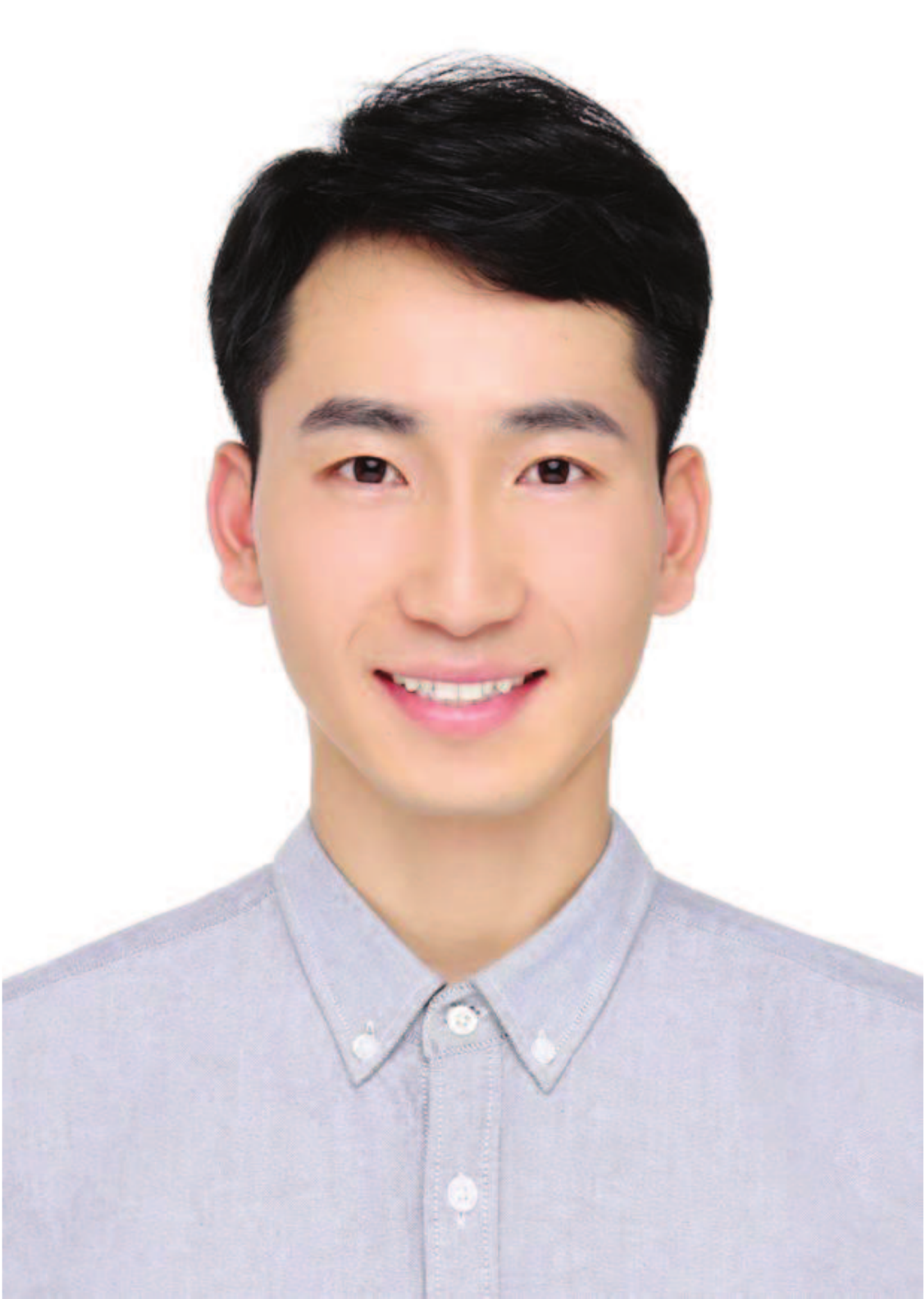}}]{Siyuan~Gu} received the B.S. degree in mathematics from Officers College of PAP, Chengdu, China, in 2015. He is currently working towards the M.S. degree in College of Systems Engineering, National University of Defense Technology, Changsha, China. His research interests include edge computing and distributed computing.\end{IEEEbiography}\vspace{-0.35in}
\begin{IEEEbiography}[{\includegraphics[width=1in,height=1.25in,clip,keepaspectratio]{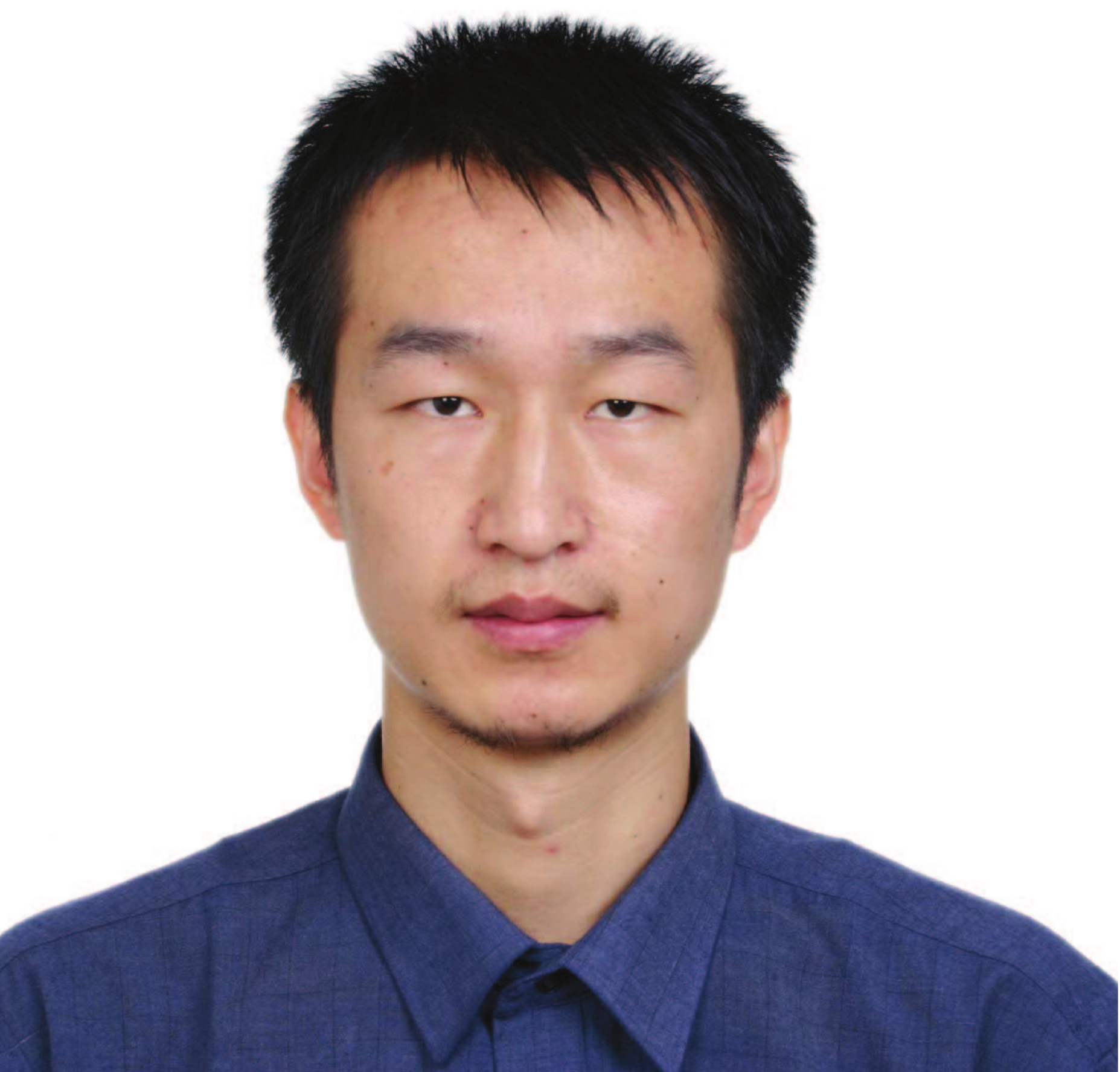}}]{Deke~Guo} received the B.S. degree in industry engineering from the Beijing University of Aeronautics and Astronautics, Beijing, China, in 2001, and the Ph.D. degree in management science and engineering from the National University of Defense Technology, Changsha, China, in 2008. He is currently a Professor with the College of System Engineering, National University of Defense Technology, and is also with the College of Intelligence and Computing, Tianjin University. His research interests include distributed systems, software-defined networking, data center networking, wireless and mobile systems, and interconnection networks. He is a senior member of the IEEE and a member of the ACM.\end{IEEEbiography}\vspace{-0.35in}
\begin{IEEEbiography}[{\includegraphics[width=1in,height=1.25in,clip,keepaspectratio]{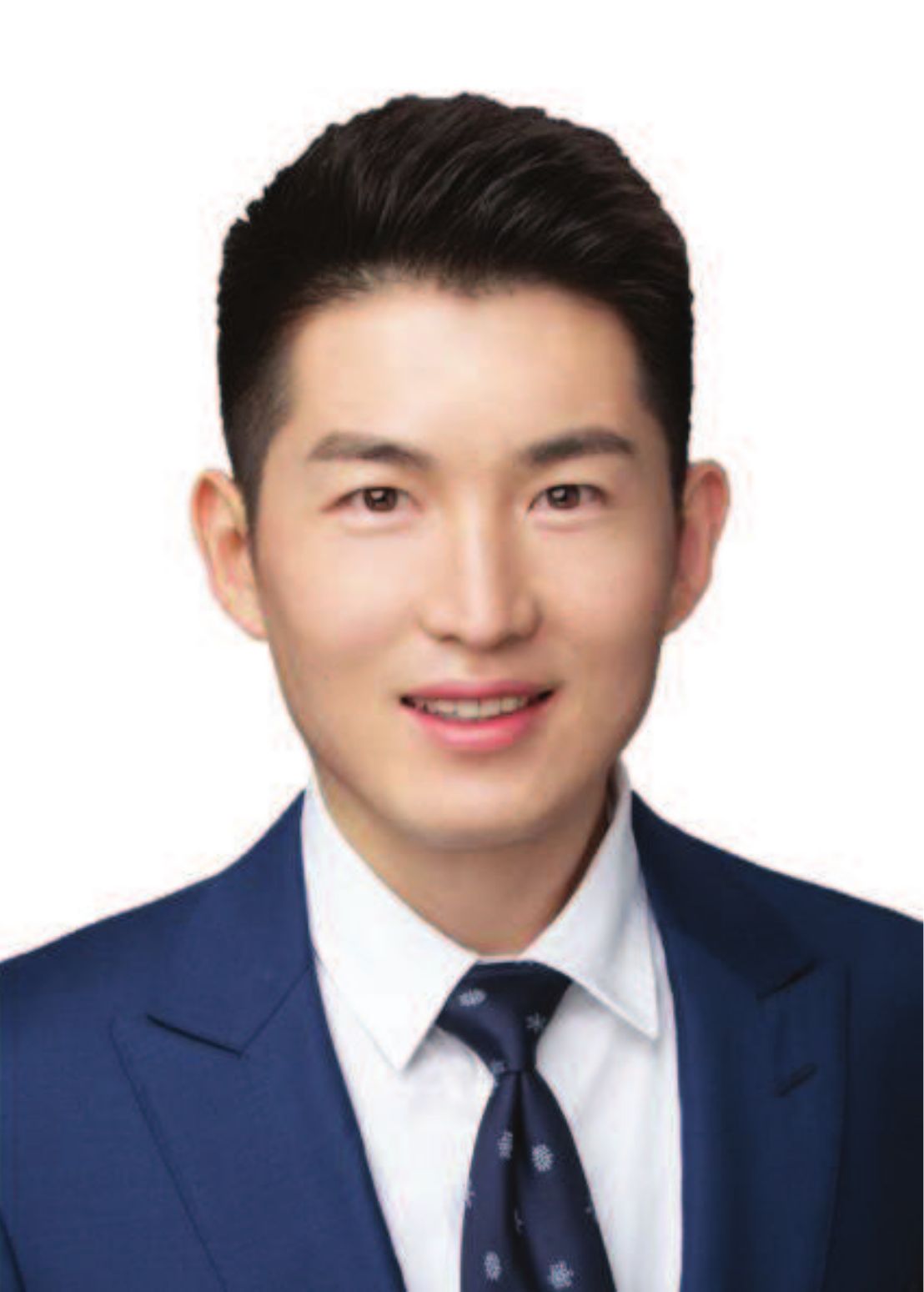}}]{Guoming~Tang} is an Assistant Professor at the Key Laboratory of Science and Technology on Information System Engineering, National University of Defense Technology (NUDT), China. He received the Ph.D. degree in Computer Science from the University of Victoria, Canada, in 2017, and both the Bachelor¡¯s and Master¡¯s degrees from NUDT in 2010 and 2012, respectively. Aided by machine learning and optimization techniques, his research focuses on computational sustainability in distributed and networking systems.\end{IEEEbiography}\vspace{-0.35in}
\begin{IEEEbiography}[{\includegraphics[width=1in,height=1.25in,clip,keepaspectratio]{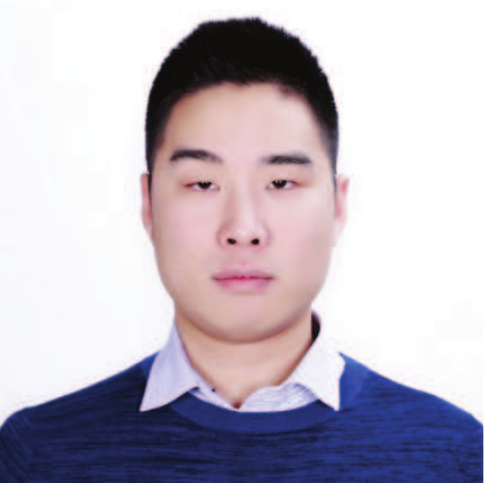}}]{Yuchen~Sun} received the B.S. degree in
Telecommunication from Huazhong University of Science and Technology, Wuhan, China, in 2018. He is currently working towards the PH.D. degree in College of Systems Engineering, National University of Defense Technology, Changsha, China. His research interests include mobile computing and machine learning.\end{IEEEbiography}\vspace{-0.35in}
\begin{IEEEbiography}[{\includegraphics[width=1in,height=1.25in,clip,keepaspectratio]{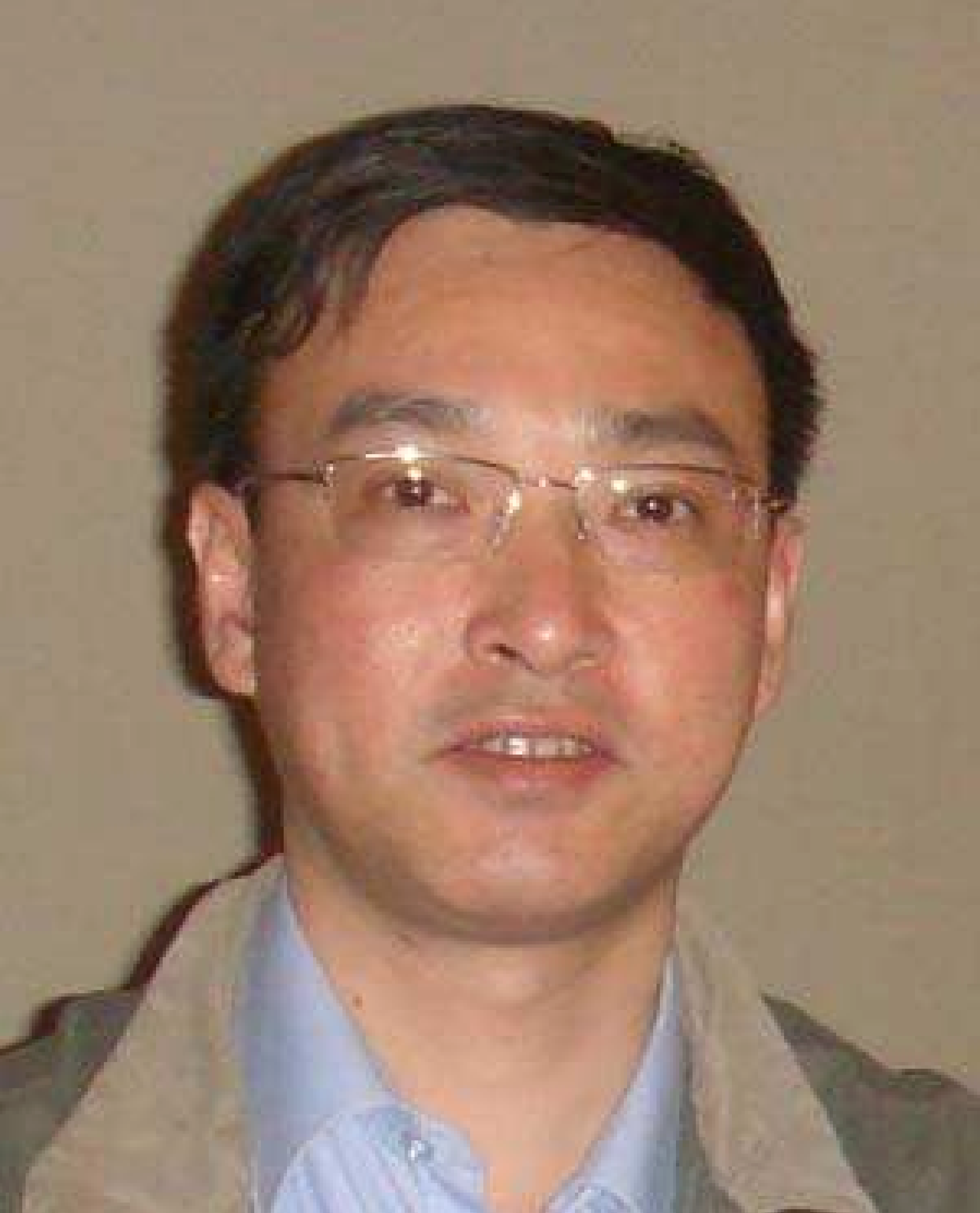}}]{Xueshan~Luo} received his B.E. degree in Information Engineering from Huazhong Institute of Technology, Wuhan, China, in 1985, and his M.S. and Ph.D degrees in System Engineering from the National  University of Defense Technology, Changsha, China, in 1988 and 1992, respectively. Currently, he is a professor of College of Systems Engineering, National University of Defense Technology. His research interests are in the general areas of information system and operation research.\end{IEEEbiography}\vspace{-0.35in}

\end{document}